%% file: SUS-16-049_temp.tex
\pdfoutput=1

\documentclass[11pt,twoside,a4paper,cmspaper,final,collab]{cms-tdr}
\def\svnVersion{428719P}\def\cmsCernNoTag{CERN-EP-2017-129}\def\cmsCernDate{\today}\def\cmsMessage{Published in the Journal of High Energy Physics as \href{http://dx.doi.org/10.1007/JHEP10(2017)005}{\doi{10.1007/JHEP10(2017)005.}}}

\begin{document}\cmsNoteHeader{SUS-16-049}

\hyphenation{had-ron-i-za-tion}
\hyphenation{cal-or-i-me-ter}
\hyphenation{de-vices}
\RCS$Revision: 423123 $
\RCS$HeadURL: svn+ssh://svn.cern.ch/reps/tdr2/papers/SUS-16-049/trunk/SUS-16-049.tex $
\RCS$Id: SUS-16-049.tex 423123 2017-08-30 22:44:08Z gouskos $

\newcommand{\dm}{\ensuremath{\Delta m}}
\newcommand{\nb}{N_{\cPqb}}
\newcommand{\nj}{N_{\text{j}}}
\newcommand{\nt}{N_{\cPqt}}
\newcommand{\nrt}{N_{\text{res}}}
\newcommand{\nw}{N_{\PW}}
\newcommand{\nsv}{N_{\text{SV}}}
\newcommand{\met}{\ptmiss}
\newcommand{\metg}{\ensuremath{p^{\text{miss},\gamma}_{\mathrm{T}}\xspace}}
\newcommand{\metll}{\ensuremath{p^{\text{miss},\ell\ell}_{\mathrm{T}}\xspace}}
\newcommand{\tfll}{TF_{\text{LL}}}
\newcommand{\znunu}{\ensuremath{\cPZ(\cPgn\cPgn)}}
\newcommand{\znunuM}{\cPZ\to\cPgn\cPgn}
\newcommand{\zll}{\ensuremath{\cPZ(\ell\ell)}}
\newcommand{\ttbarW}{\ensuremath{\ttbar\PW}}
\newcommand{\ttbarZ}{\ensuremath{\ttbar\cPZ}}
\newcommand{\ttZ}{\ttbarZ}
\newcommand{\ttW}{\ttbarW}
\newcommand{\ww}{\PW\PW}
\newcommand{\wz}{\PW\cPZ}
\newcommand{\zz}{\cPZ\cPZ}
\newcommand{\mtb}{m_{\mathrm{T}}^{\mathrm{b}}}
\newcommand{\htt}{H_{\mathrm{T}}}
\newcommand{\metsig}{\ensuremath{S_{E_{\mathrm{T}}\hspace{-0.8em}/\kern0.45em}\xspace}}
\newcommand{\dphijonetwothree}{\Delta\phi_{123}}
\newcommand{\dphijonetwothreefour}{\Delta\phi_{1234}}
\newcommand{\ptisr}{p_{\mathrm{T}}^{\mathrm{ISR}}}
\newcommand{\ptb}{\ensuremath{\pt^{\mathrm{b}}}}
\newcommand{\ptbonetwo}{\ptb}
\newcommand{\stopq}{\PSQt_{1}}
\newcommand{\stopqonetwo}{\PSQt}
\newcommand{\stopqbar}{\PASQt_{1}}
\newcommand{\stopqonetwobar}{\PASQt}
\newcommand{\sbottomqonetwo}{\PSQb}
\newcommand{\topq}{\cPqt}
\newcommand{\topqbar}{\cPaqt}
\newcommand{\bq}{\cPqb}
\newcommand{\cq}{\cPqc}
\newcommand{\lsp}{\PSGczDo}
\newcommand{\lspall}{\PSGcz}
\newcommand{\chg}{\PSGcpm}
\newcommand{\chgp}{\PSGcpDo}
\newcommand{\chipmone}{\PSGcpmDo}
\newcommand{\rjet}{r_{\text{jet}}}
\newcommand{\prjet}{r^{\mathrm{pseudo}}_{\text{jet}}}
\newcommand{\MT}{\ensuremath{m_{\mathrm{T}}}\xspace}
\newcommand{\wjets}{\PW+jets\xspace}
\newcommand{\zjets}{\cPZ+jets\xspace}
\newcommand{\gjets}{\ensuremath{\gamma}+jets\xspace}
\newcommand{\ljets}{lepton+jets\xspace}

\cmsNoteHeader{SUS-16-049}
 \title{Search for direct production of supersymmetric partners of the top quark in the all-jets final state in proton-proton collisions at $\sqrt{s} = 13\TeV$}

\date{\today}

\abstract{A search for direct production of top squark pairs in events with jets and large transverse momentum imbalance is presented. The data are based on proton-proton collisions at a center-of-mass energy of 13\TeV, collected with the CMS detector in 2016 at the CERN LHC, and correspond to an integrated luminosity of 35.9\fbinv. The search considers a variety of $R$-parity conserving supersymmetric models, including ones for which the top squark and neutralino masses are nearly degenerate. Specialized jet reconstruction tools are developed to exploit the unique characteristics of the signal topologies. With no significant excess of events observed above the standard model expectations, upper limits are set on the direct top squark pair production cross section in the context of simplified supersymmetric models for various decay hypotheses. Models with larger differences in mass between the top squark and neutralino are probed for masses up to 1040 and 500\GeV, respectively, whereas models with a more compressed mass hierarchy are probed up to 660 and 610\GeV, respectively. The smallest mass difference probed is for masses near to 550 and 540\GeV, respectively.}

\hypersetup{%
pdfauthor={CMS Collaboration},%
pdftitle={Search for direct production of supersymmetric partners of the top quark in the all-jets final state in proton-proton collisions at sqrt(s) = 13 TeV},%
pdfsubject={CMS},%
pdfkeywords={CMS, SUSY, stop}}

\maketitle
 \section{Introduction}
\label{sec:intro}
Although the standard model (SM) of particle physics provides a remarkably accurate description of phenomena associated with the known elementary particles and their interactions, it leaves significant problems unresolved. It cannot, for instance, explain how the Higgs boson \cite{Aad:2012tfa,Chatrchyan:2012ufa,Chatrchyan:2013lba,Aad:2014aba,Khachatryan:2014jba,Aad:2015zhl} can evade divergent quantum corrections, without very significant fine tuning~\cite{Wilson:1970ag,Giudice:2013nak} of SM parameters, to allow it to have its mass at the weak scale \cite{'tHooft:1979bh,Witten:1981nf,Dine:1981za,Dimopoulos:1981au,Dimopoulos:1981zb,Kaul:1981hi}. Moreover, an abundance of cosmological observations, including the existence of dark matter, cannot be explained within the context of the SM alone~\cite{Zwicky:1933gu,Rubin:1970zza,Olive:2016xmw}.

Supersymmetry (SUSY) provides a theoretical framework that can address these questions. At its core, SUSY is a symmetry between fermions and bosons. In SUSY, a ``sparticle" (generally referred to as a superpartner) is proposed for each SM particle with the same gauge quantum numbers but differing by one half-unit of spin and potentially in mass. The superpartners of the electroweak vector W and Z bosons and scalar Higgs boson mix to produce charged and neutral fermions referred to as charginos (\chg) and neutralinos (\lspall), respectively. For a given fermion f, there are two superpartners corresponding to the fermion's left- and right-handed states. The superpartners mix to form two mass eigenstates, $\widetilde{\text{f}}_1$ and $\widetilde{\text{f}}_2$, with $\widetilde{\text{f}}_1$ being the lighter of the two. The quantum corrections to the value of the Higgs boson mass ($m_\mathrm{H}$) from sparticles could cancel the otherwise problematic SM contributions. In this way, SUSY can protect the value of $m_\mathrm{H}$ \cite{Barbieri:1987fn,deCarlos:1993rbr,Dimopoulos:1995mi,Barbieri:1995uv}, provided that the mass differences between the SM particles and their superpartners are not too large. This is particularly important for superpartners of third generation SM particles, because they have the largest couplings to the Higgs boson, and therefore produce the largest corrections. Furthermore, a combination of precision measurements and null search results indicate that the superpartners of the light quarks may have very large masses \cite{Feng:2013pwa}. In view of these considerations, the superpartners of the top and bottom quarks, the $\stopqonetwo$ and $\sbottomqonetwo$ squarks, respectively, are expected to be among the lightest sparticles, potentially light enough to be produced at the CERN LHC \cite{LHC}. An important point to note is that SUSY models with $R$-parity conservation \cite{Wess:1974tw,Farrar:1978xj} require sparticles to be produced in pairs, with the lightest SUSY particle (LSP) therefore stable on cosmological time scales. This means that if the lightest neutralino, denoted \lsp, is the LSP, then it is also a very promising dark matter candidate \cite{Feng:2010gw} that would remain at the end of all $R$-parity conserving sparticle cascade decays. The two motivating principles above place the search for pair production of top squarks ($\stopqonetwo\stopqonetwobar$) among the highest priorities of the LHC program.

The most recent searches for direct $\stopqonetwo\stopqonetwobar$ production were carried out by the ATLAS and CMS Collaborations in proton-proton (\Pp\Pp) collisions at center-of-mass energies $\sqrt{s}$ of 7, 8, and 13\TeV~at the LHC \cite{ATLAS1,ATLAS2,ATLAS5,ATLAS5a,ATLAS6,ATLAS7,ATLAS8,Aad:2015pfx,CMS-STOP-lepton,CMS-alphaT,RAZOR_8TeV,stop8TeV,Khachatryan:2016oia,Aaboud:2016lwz,Khachatryan:2016xvy,Khachatryan:2016epu,Khachatryan:2016dvc,Sirunyan:2016jpr,Aaboud:2017ejf,Sirunyan:2017cwe,Sirunyan:2017kqq}.
 The searches have provided no evidence for sparticle production in models with $\stopqonetwo$~masses up to ${\sim}900\GeV$ and \lsp~masses up to ${\sim}400\GeV$.

\begin{figure}[!pt]
\centering
\includegraphics[width=0.95\textwidth]{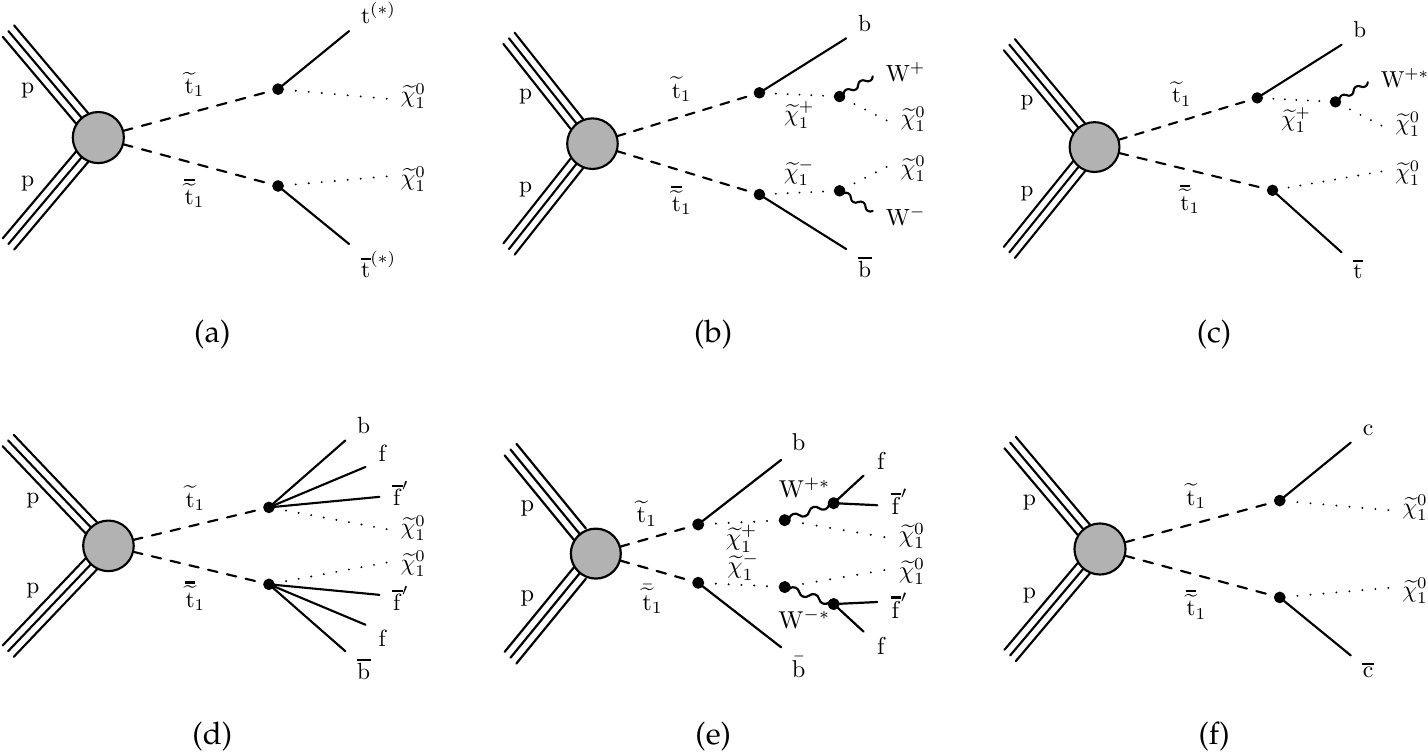}
\caption{
\label{fig:diagram} Diagrams for the decay modes of pair-produced top squarks studied in this analysis. The decay cascades are denoted: (a) T2tt, (b) T2bW, (c) T2tb, (d) T2ttC, (e) T2bWC, and (f) T2cc. An asterisk indicates that the particle may be produced off-shell.}
\end{figure}

This paper presents a search for direct $\stopqonetwo\stopqonetwobar$ production in $R$-parity conserving SUSY using data collected in \Pp\Pp~collisions at $\sqrt{s} =  13\TeV$ by the CMS experiment at the LHC in 2016, and corresponding to an integrated luminosity of 35.9\fbinv. The search is based on methods presented in Ref.~\cite{Sirunyan:2016jpr}, and represents an extension of that search to larger sparticle masses by means of a significantly larger dataset and the development of more sensitive search tools. This search focuses on all-hadronic final states, defined as those events whose visible content is made up solely of hadronic jets, as would be expected for signal processes in which all \PW~bosons decay to quarks. These final states have the largest accessible branching fraction. In many SUSY models, the favored $\stopqonetwo$ decay modes depend strongly on the mass hierarchy of the sparticles.  In particular, different ranges of mass difference \dm~between the $\stopqonetwo$ and $\lsp$ correspond to very different final-state signatures. Only the lightest $\stopqonetwo$ mass eigenstate, $\stopq$, is assumed to be involved in the models considered in this paper, although the results are expected to be equivalent for the heavier eigenstate. The $\stopq$ decay modes of the simplified models~\cite{Alwall:2008ag,Alwall:2008va,Alves:2011wf} that are used as the basis for our searches are displayed in Fig.~\ref{fig:diagram}.

The search regions (SR) are optimized for different models and ranges of \dm. In models with \dm~larger than the \PW~boson mass $m_{\text{W}}$ (``high \dm~models''), the simplest decays that we consider are $\stopq \to \topq^{(*)} \lsp$, denoted ``T2tt", and $\stopq \to \bq \chipmone \to \bq \PW^{\pm} \lsp$, denoted ``T2bW", under the assumption that the $\chipmone$ mass lies halfway between the $\stopq$ and $\lsp$ masses. The choice of moderate $\chipmone$ mass in the latter model permits high momentum objects in the final state. The $\chipmone$ represents the lightest chargino, and $\lsp$ is the stable LSP, which escapes detection to produce a large transverse momentum imbalance in the event. Another model, denoted ``T2tb", is considered under the assumption of equal branching fractions of the two aforementioned decay modes. This model, however, assumes a compressed mass spectrum in which the mass of the $\chipmone$~is only 5\GeV greater than that of the \lsp. As a result, the \PW~bosons from chargino decays are produced far off-shell.

In models with \dm~less than $m_{\text{W}}$ (``low \dm~models''), the $\stopq$ can decay through the T2tt decay mode with off-shell \topq~and \PW, through the same decay chain as in the T2bW model, via off-shell \PW~bosons, or decay through a flavor changing neutral current process ($\stopq \to \cq\lsp$, where $\cq$ is the charm quark). These will be referred to as the ``T2ttC'', ``T2bWC'', and ``T2cc'' models, respectively, where C denotes the hypothesis of a compressed mass spectrum in the first two cases.
Observations in such low \dm~models are experimentally challenging since the visible decay products are typically very soft (low-momentum), and therefore often evade identification. Nevertheless, such models are particularly interesting because their dark matter relic density is predicted to be consistent with the cosmological observations~\cite{Balazs:2004bu}.  Specialized jet reconstruction tools and event selection criteria are therefore developed to enhance sensitivity to these signals.

This paper is organized as follows. A brief description of the CMS detector is presented in Section 2, while Section 3 discusses the simulation of background and signal processes. Event reconstruction is presented in Section 4, followed by a description of the search strategy in Section 5. Methods employed to estimate the SM backgrounds and their corresponding systematic uncertainties are detailed in Sections 6 and 7, respectively. The discussion of the systematic uncertainties assigned to the signal processes is also presented in Section 7. The results of the search and their interpretation in the context of a variety of models of $\stopq$ production and decay are presented in detail in Section 8, followed by a summary in Section 9.

\section{The CMS detector}
The central feature of the CMS apparatus is a superconducting solenoid of 6\unit{m} internal diameter, providing a magnetic field of 3.8\unit{T}. Within the solenoid volume are an all-silicon pixel and strip tracker, a lead tungstate crystal electromagnetic calorimeter (ECAL), and a brass and scintillator hadron calorimeter (HCAL), each composed of a barrel and two endcap sections. Forward calorimeters extend the pseudorapidity ($\eta$) coverage provided by the barrel and endcap detectors. Muons are measured in gas-ionization detectors embedded in the steel flux-return yoke outside the solenoid. The first level of the CMS trigger system, composed of custom hardware processors, uses information from the calorimeters and muon detectors to select the most interesting events in a fixed time interval of less than 4\mus. The high-level trigger processor farm further decreases the event rate from around 100\unit{kHz} to around 1\unit{kHz}, before data storage. A more detailed description of the CMS detector, together with a definition of the coordinate system used and the relevant kinematic variables, can be found in Ref.~\cite{JINST}.

\section{Simulated events}

Monte Carlo (MC) simulated events are used to study the important SM backgrounds, as well as to formulate the overall search for SUSY processes. Background processes composed uniquely of jets produced via the strong interaction of quantum chromodynamics (QCD) are referred to as ``QCD multijet" processes. Simulated events originating from \ttbar, \PW+jets, $\Z$+jets, $\gamma$+jets, and QCD multijet processes are generated using \MGvATNLO 2.3.3~\cite{Alwall:2014hca} at leading order (LO) using the LO NNPDF3.0~\cite{Ball:2014uwa} parton distribution functions (PDF). The \wz, \zz, \ttZ, and \ttW~processes are generated using $\MADGRAPH5\textsc{\_aMC@NLO}$ at next-to-leading order (NLO), the single top quark process in the \cPqt\PW~channel using \POWHEG~\cite{Nason:2004rx,Frixione:2007vw,Alioli:2010xd,Re:2010bp} and the \ww~process is generated at NLO with \POWHEG v2.0~\cite{Nason:2013ydw}, all using the NLO NNPDF3.0 PDF. In all of the aforementioned cases,  parton showering and hadronization are simulated in \PYTHIA8.212~\cite{pythia8}. The potential for double counting of partons generated using \PYTHIA with those using \MGvATNLO is minimized using the MLM~\cite{Alwall:2007fs} and the FXFX~\cite{Frederix:2012ps} matching schemes, in the LO and NLO samples, respectively. To evaluate systematic uncertainties associated with these aspects of event simulation, two additional \ttbar~samples are generated using \POWHEG~v2.0~\cite{Campbell:2014kua}, where one is interfaced with \PYTHIA~and the other with \HERWIGpp v2.7.1~\cite{Bahr:2008pv}. Additional QCD multijet samples are also generated, but interfaced with~\HERWIGpp~for the modeling of parton showering and hadronization. Signal processes are generated at LO using \MGvATNLO based on the LO NNPDF3.0 PDF with \PYTHIA~used for parton showering and hadronization. Signal production cross sections are calculated using NLO with next-to-leading logarithm (NLL) soft-gluon resummations (NLO+NLL)~\cite{Borschensky:2014cia}. The most precise cross section calculations are used to normalize the SM simulated samples, corresponding to next-to-next-to-leading order (NNLO) accuracy~\cite{Czakon:2011xx,Li:2012wna,Kant:2014oha,Kidonakis:2012rm} in most cases. Finally, the transverse momentum (\ptvec, with magnitude \pt) spectrum of top quarks in \ttbar~events is reweighted (referred to as ``top quark \pt~reweighting'') to account for effects due to missing higher-order corrections in MC simulation, according to the results presented in Ref.~\cite{Khachatryan:2015fwh}.

 A full \GEANTfour-based model~\cite{geant4} is used to simulate the response of the CMS detector to SM background samples. The CMS fast simulation package~\cite{fastsim} is used for signal samples after verifying  that it provides results that are consistent with those obtained from the full \GEANTfour-based simulation. Event reconstruction is treated in the same manner for MC simulation as for data. A nominal distribution of multiple \Pp\Pp~collisions in the same or neighboring bunch crossings (referred to as ``pileup'') is used to overlay the simulated events. The events are then reweighted to match the pileup profile observed in the collected data.

\section{Event reconstruction}
\label{sec:reco}

Events are reconstructed using the CMS particle-flow (PF) algorithm~\cite{Sirunyan:2017ulk}, which combines information from all detector subsystems to reconstruct the properties of the final-state particles produced in the \Pp\Pp~collisions. At least one reconstructed vertex is required; for multiple collision vertices from pileup interactions, the reconstructed vertex with the largest value of summed physics-object $\pt^2$ is taken to be the primary $\Pp\Pp$ interaction vertex (PV). The physics objects used in this context are the objects returned by a jet finding algorithm~\cite{Cacciari:2008gp,Cacciari:2011ma} applied to all charged tracks associated with the vertex under consideration, plus the corresponding associated missing transverse momentum (the precise definition is given later in the text). Events affected by instrumental noise or reconstruction failures are identified through dedicated filters and rejected. Reconstructed particles are identified as charged hadrons, neutral hadrons, electrons, muons, or photons, to constitute a list of PF candidates.

Our primary jet collection is produced by clustering the PF candidates originating from the PV using the anti-\kt~algorithm~\cite{Cacciari:2008gp} with a distance parameter of 0.4. The jet energy is corrected for the contribution from pileup based on the jet area method~\cite{Cacciari:2007fd,CMS-PAS-JME-14-001}. Additional corrections to the jet energy scale are applied to compensate for nonuniform detector response~\cite{Khachatryan:2016kdb}. Jets are required to have $\pt \geq 20\GeV$ and be contained within the tracker volume of $\abs{\eta} \leq 2.4$.

Jets originating from the hadronization of bottom (\bq) quarks are identified, or ``tagged", through the combined secondary vertex (CSVv2) \bq~tagging algorithm~\cite{Chatrchyan:2012jua,CMS-PAS-BTV-15-001}. The working point used provides an efficiency for the \bq~tagging of jets originating from \bq~quarks that varies from 60 to 75\% depending on $\pt$, whereas the misidentification rate for light quarks or gluons is ${\sim}1\%$, and ${\sim}15\%$ for charm quarks. A novel soft b tagging algorithm was developed for this analysis and used to identify \bq~quarks with $\ptb < 20\GeV$ (i.e. below the jet $\pt$ threshold). The algorithm is described in Section~\ref{sec:softbtag}. Although the T2cc model involves charm quark jets in the final state, no dedicated \cq~tagger was used in this analysis.

To estimate the $\pt$ imbalance in the event, the missing transverse momentum, \ptvecmiss, is defined as the negative of the vectorial sum of the \ptvec~of all PF candidates in the event. Its magnitude is denoted \met. The jet energy scale corrections applied to the jets are propagated to \ptvecmiss.

Electrons are reconstructed by combining information from the inner tracker with energy depositions in the ECAL ~\cite{Khachatryan:2015hwa}. Muons are reconstructed by combining tracks in the inner tracker and in the muon system ~\cite{Chatrchyan:2012xi}. Tracks associated with electrons or muons are required to originate from the PV, and a set of quality criteria is imposed to assure efficient identification ~\cite{Khachatryan:2015hwa,Chatrchyan:2012xi}. To suppress misidentification of charged hadrons as leptons, we require electrons and muons to be isolated from jet activity within a \pt-dependent cone size defined by a radius $R_\text{rel}$ in the $\eta$-$\phi$ plane, where $\phi$ is the azimuthal angle in radians. The relative isolation, $I_\text{rel}$, is defined as the scalar sum of the $\pt$ of the PF candidates  within the cone divided by the lepton \pt. Charged PF candidates not originating from the PV, as well as PF candidates identified as electrons or muons, are not considered in the sum.  The cone size $R_\text{rel}$ depends on the lepton \pt:

\begin{equation}
R_\text{rel} =
\begin{cases}
0.2, & \pt < 50\GeV, \\
10\GeV/\pt, & 50 \leq \pt < 200\GeV, \\
0.05, & \pt \geq 200\GeV.
\end{cases}
\end{equation}

The decreasing cone radius at larger \pt provides high efficiency for the collimated decay products of highly Lorentz-boosted heavy objects~\cite{Rehermann:2010vq}. The isolation sum $I_\text{rel}$ is corrected for contributions of neutral particles originated from pileup interactions using an area-based estimate~\cite{CMS-PAS-JME-14-001} of pileup energy deposition in the cone.

Photons are reconstructed from energy depositions in the ECAL using identification algorithms that utilize a collection of variables related to the spatial distribution of shower energy in the supercluster (a group of 5x5 ECAL crystals), the photon isolation, and the fraction of the energy deposited in the HCAL behind the supercluster relative to the energy observed in the supercluster~\cite{Khachatryan:2015iwa}.

Tau lepton decays to hadrons, $\tauh \to (\textrm{hadrons}) \nu_{\tau}$, are reconstructed starting from isolated charged-hadron candidates with $\pt \geq 10$\GeV~and $\abs{\eta} \leq 2.4$. If there are photons with $\pt \geq 0.5\GeV$ within a cone of $R \leq 0.2$ around the charged hadron, the leading \pt~photon momentum is vectorially added to that of the charged hadron candidate. In addition, we impose a requirement on the transverse mass of the $\tauh$; for an object with transverse momentum \ptvec, the transverse mass \MT is defined as:

\begin{equation}
m_\mathrm{T}({\vec{p}_{\mathrm{T}}}, \ptvecmiss) = \sqrt{2 p_{\mathrm{T}} \met (1 - \cos \Delta\phi)} ~~,\label{eq:MT}
\end{equation}

where $\Delta\phi$ is the difference in azimuthal angle between \ptvec and \ptvecmiss. We require the transverse mass of the $\PGt_{\rm h}$ to be less than 100\GeV, consistent with the expectation from a $\tauh$ emitted in a \PW~boson decay in a high-multiplicity jet environment. A multivariate boosted decision tree (BDT) classifier~\cite{Roe:2004na} is trained to distinguish $\tauh$ decay products from other charged hadrons. Input variables include isolation sums within cones of several radii, $R$-distances from the $\tauh$ candidate to the nearest charged particle and to the axis of the jet in which it is contained, and the \bq~tagging discriminant value of that jet.

Many of the $\stopq$~decay modes involve unique final-state signatures. In view of this, reconstruction tools have been developed to exploit these signatures while significantly suppressing the SM background. Signal models with large \dm~have decay chains involving on-shell top quarks and \PW~bosons. Identification of jets associated with the decays of top quarks and \PW~bosons to quarks is an important component of the analysis, used to suppress most of the backgrounds in searches that target such signals. Because they exhibit a wide range of Lorentz boosts, we take different approaches in their reconstruction depending on whether they have large or small \pt; these are described in Sections~\ref{sec:boostedtopw} and~\ref{sec:resolvedtop}, respectively. In contrast, the decay products in models with small \dm~are very soft and often fail to be reconstructed through the standard algorithms. We have therefore developed more effective algorithms for these cases that are described in Sections~\ref{sec:isrjets} and~\ref{sec:softbtag}.

\subsection{\texorpdfstring{Identification of high-\pt top quarks and \PW~bosons}{Identification of high-pt top quarks and W bosons}}
\label{sec:boostedtopw}
The decay products of highly boosted top quarks with $\pt \geq 400$\GeV, or \PW~bosons with $\pt \geq 200$\GeV, are usually contained within a cone of radius $R = 0.8$~\cite{Shelton:2013an}. A collection of ``large-$R$ jets'', which is distinct from, and possibly overlaps with, the collection of primary jets, is used to reconstruct these boosted objects by means of the anti-\kt~clustering algorithm with a distance parameter of 0.8. Additional information on jet substructure is obtained by reclustering the constituents of these jets through the Cambridge--Aachen algorithm~\cite{Dokshitzer:1997in}. The ``modified mass drop tagger'' algorithm ~\cite{Dasgupta:2013ihk}, also known as the ``soft drop'' (SD) algorithm, with angular exponent $\beta = 0$, soft cutoff threshold $z_{\mathrm{cut}} \leq 0.1$, and characteristic radius $R_{0} = 0.8$~\cite{Larkoski:2014wba}, is applied to remove soft, wide-angle radiation from the jet. The performance of the SD algorithm does not depend on the algorithm used initially to reconstruct the large-$R$ jets. Top quark and \PW~boson candidates are selected from the collection of large-$R$ jets after applying a loose preselection based on variables reconstructed using the SD algorithm. In our configuration, the SD algorithm identifies two hard subjets of the large-$R$ jet by reversing the Cambridge--Aachen clustering history. The two hard substructures should correspond to the \PW~boson and b quark jet, in the case of top quark candidates, or to two quark jets of a \PW~boson decay, in the case of a \PW~boson candidate. The top quark (\PW~boson) candidates are required to have soft-drop mass m$_{\text{SD}} \geq 110$~($50 \leq \text{m}_{\text{SD}} < 110$)\GeV, $\pt \geq 400 (200)$\GeV, $\abs{\eta} \leq 2.4$, and subjets with \pt$ \geq 20$\GeV. These m$_{\text{SD}}$ requirements incur minimal efficiency losses, and ensure that candidates can only be tagged uniquely.

\begin{figure}[ht]
\centering
\includegraphics[width=0.45\textwidth]{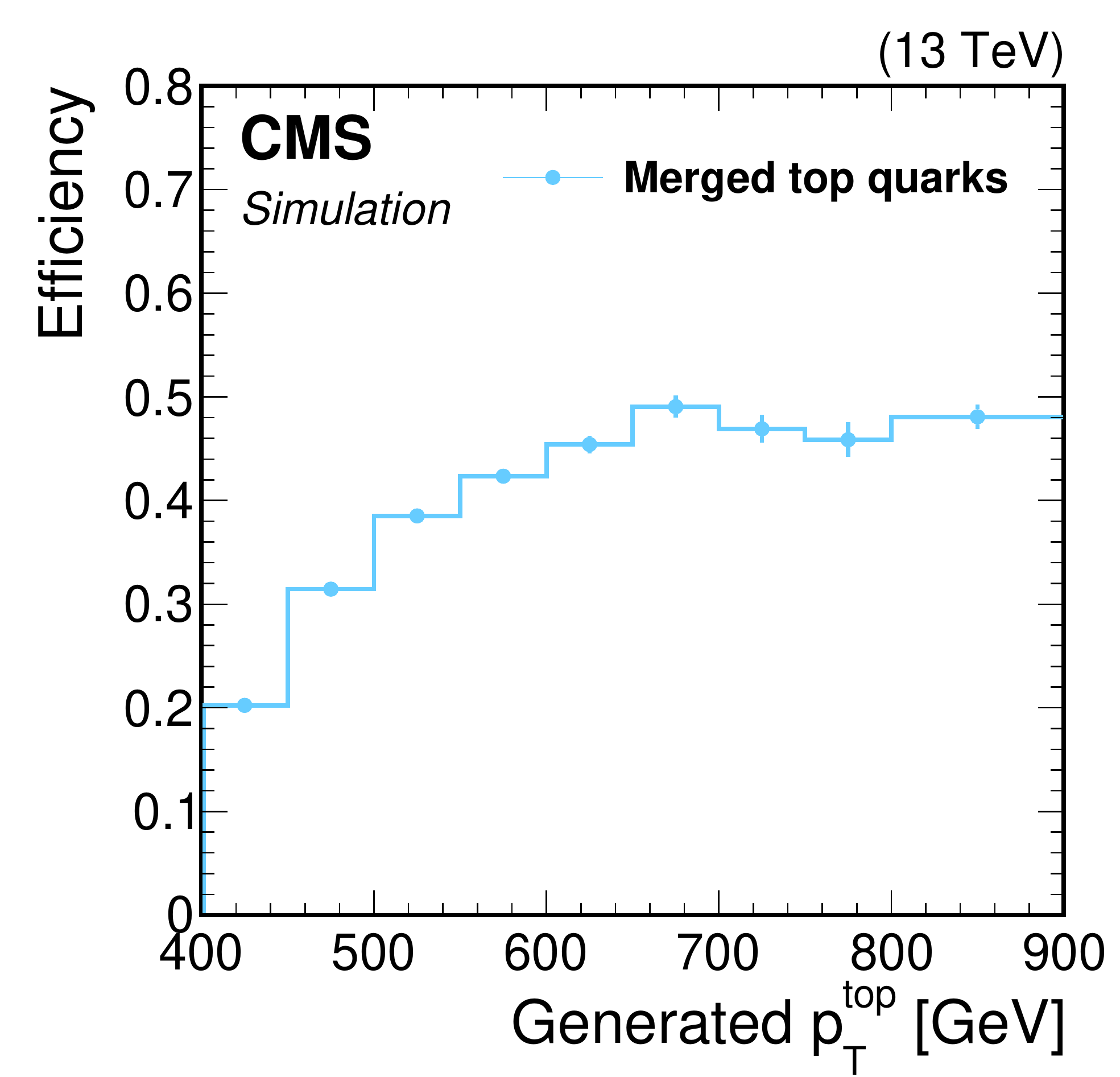}
\includegraphics[width=0.45\textwidth]{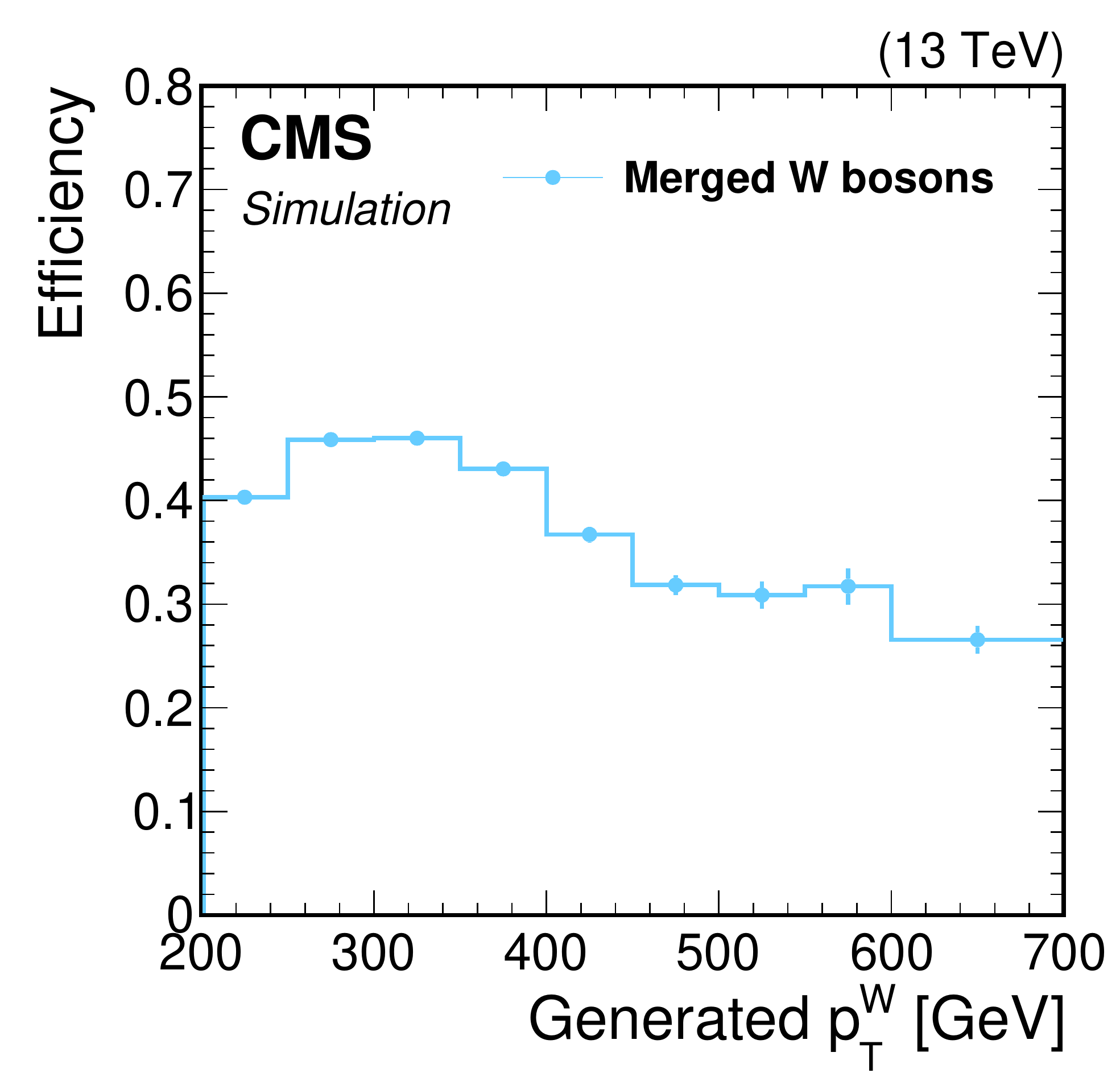}

\caption{Efficiencies in MC simulation for identifying the quark decays of top quarks (left), and \PW~bosons (right), as a function of the \pt~of the generated top quarks or \PW~bosons to which they were matched.}
\label{fig:boostedtopweff}
\end{figure}

Two separate multivariate BDT are trained to identify candidates for the quark decays of highly boosted top quarks and \PW~bosons. The identified objects are subsequently referred to as ``merged" top quarks and \PW~bosons, respectively. The input variables to the two BDT rely on m$_{\text{SD}}$, N-subjettiness ratios ($\tau_{\rm3}/\tau_{\rm2}$ and $\tau_{\rm2}/\tau_{\rm1}$)~\cite{nsubjettiness}, observables related to quark-gluon discrimination~\cite{CMS-PAS-JME-13-002}, the \bq~tagging discriminant value, the relative difference in \pt~between each of the two subjets within the large-$R$ jet, and the mass of each subjet. The N-subjettiness variable, $\tau_{N}$, is a measure of the degree to which the jet originates from N subjets. The BDT are trained in MC simulated samples using the Toolkit for Multivariate Data Analysis (TMVA)~\cite{Hocker:2007ht} to discriminate between ``background" and ``signal" large-$R$ jets. The merged top quark BDT is trained using, for ``signal", candidates that are matched to generated quark decays of top quarks in simulated SUSY events. For the merged \PW~boson BDT  this procedure is repeated in simulated \ttbar events. For the ``background" we consider the remaining candidates that could not be matched.
The efficiencies to identify matched top quarks and \PW~bosons are shown in Fig.~\ref{fig:boostedtopweff}. The merged \PW~boson tagging efficiency is determined using \PW~bosons originating from generated top quark decays; thus, the moderate drop at large \pt~can be largely attributed to the merging  of the top quark decay products, which reduces the effectiveness of the jet substructure variables. The misidentification rate for jets initiated by either gluons or light quarks depends on the \pt~of the large-$R$ jet and ranges from 1 to 4\% and from 2  to 10\% for merged top quarks and \PW~bosons, respectively.

The misidentification rates for these top quark and \PW~boson taggers are measured in data using a sample of multijet events that is dominated by the QCD multijet process, selected with an $\htt$~trigger, where $\htt$~is defined as the scalar sum of the \pt of the primary jets in the event. We require the events to contain at least one large-$R$ jet and $\htt \geq 1\TeV$. The misidentification rate is measured as a function of the jet $\pt$~and~$\eta$, and then compared to the expected rates in simulation. Data-to-simulation ratios are found to deviate from unity by no more than 20\%, and are used to correct results obtained with simulated event samples.

The top quark and \PW~boson tagging efficiencies are measured in data using a sample of \ljets~events dominated by the semileptonic $\ttbar$~process and selected using a single-muon trigger. The muon is required to have $\pt \geq 50\GeV$ and $\abs{\eta} \leq 2.1$. To suppress other backgrounds, at least one \bq-tagged jet is required in the same hemisphere as the muon, and the large-$R$ jet is required to be in the opposite hemisphere. Contributions from processes with no quark decays of top quarks or \PW~bosons are corrected through misidentification correction factors applied before obtaining the tagging efficiencies. These observed efficiencies are compared to those estimated in simulation, and simulation-to-data correction factors, typically ranging from 0.9 to 1.1, are extracted and applied to simulated events to account for any dependence on $\pt$. Simulated signal events generated in the CMS fast simulation package are corrected in a similar way for the differences in tagging performance relative to the full \GEANTfour-based simulations.

\subsection{Identification of intermediate-\texorpdfstring{\pt}{pt} top quarks}
\label{sec:resolvedtop}
The decay products of moderately boosted top quarks are often resolved as three separate jets in the primary jet collection. To avoid overlap with merged top quarks and \PW~bosons,
we only consider a ``cleaned" subset of jets that are separated by a distance $R > 0.8$ from all of the candidate merged objects. Three-jet ``resolved" top quark candidates are formed by starting with a jet from the cleaned jet collection that is designated to be the \bq~constituent jet. The two jets with highest \bq~tagging discriminant values are the only eligible jets for this step.  Two additional constituent jets are designated \PW~constituent jets after being identified from all two-jet combinations in the cleaned collection, excluding the already designated \bq~jet. The algorithm is repeated with the remaining \bq~jet. To reduce the combinatorial background before making any stringent selections, we require the two \PW~constituent jets to have invariant mass within $40\GeV$ of $m_{\rm W}=80\GeV$ and the combined three-jet system to have invariant mass within 80\GeV of the top quark mass $m_{\rm t}=175\GeV$. The three-jet systems that pass these requirements are considered for possible tagging as resolved top quarks.

\begin{figure}[ht]
\centering
\includegraphics[width=0.45\textwidth]{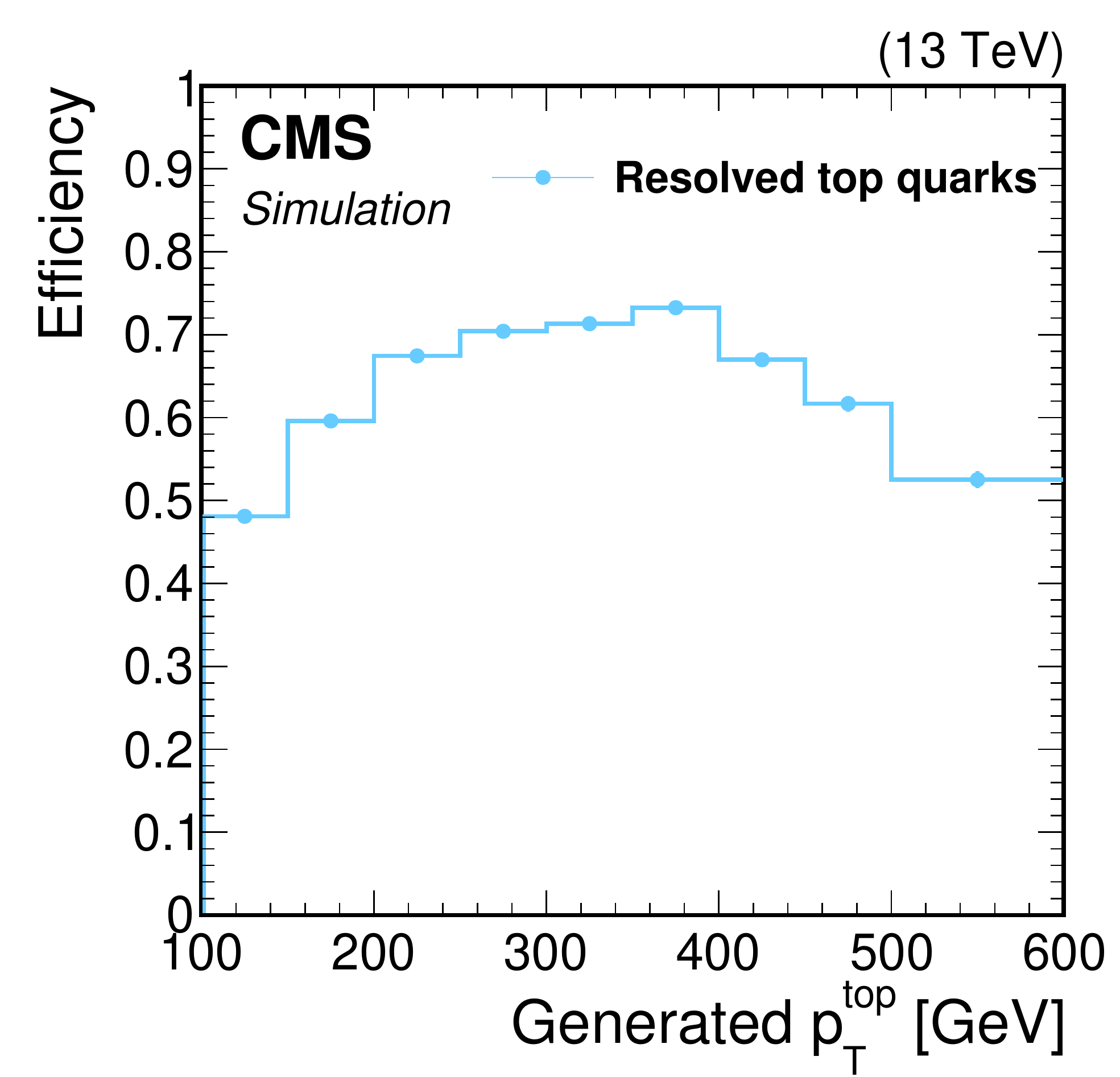}
\includegraphics[width=0.45\textwidth]{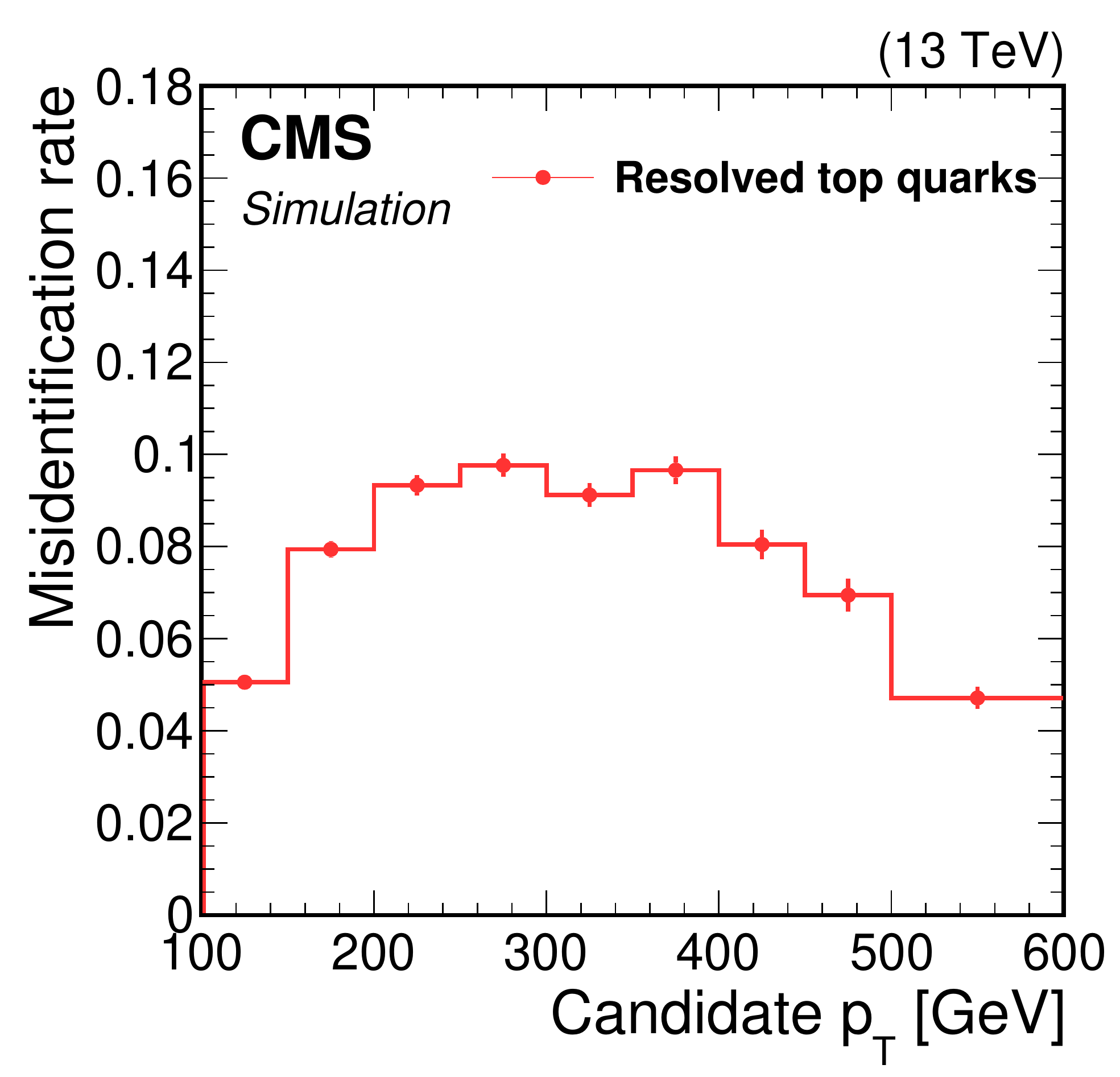}

\caption{Left: Efficiency in MC simulation to identify resolved top quark decays as a function of the \pt~of the generated top quark. Right: Misidentification rate in MC simulation as a function of the \pt~of resolved top quarks, in a sample dominated by the QCD multijet process.}
\label{fig:resolvedtopeff}
\end{figure}

Resolved top quark tagging is carried out using a BDT trained on simulated \ttbar events. It exploits properties of each three-jet candidate, including masses, angular separations, and other kinematic properties of the constituents. Additional input variables are quark-gluon discrimination metrics~\cite{CMS-PAS-JME-16-003},  \bq~tagging discriminant values, and charm quark versus light quark jet discrimination~\cite{cvsl} for each of the three jets. The performance of the resolved top quark tagger is shown in Fig.~\ref{fig:resolvedtopeff}. The drop in efficiency at very high $\pt$~stems from the fact that top quark decay products are kinematically more likely to be merged into single large-$R$ jets. Correspondingly, the efficiency of the merged top quark tagger starts to become significant in this region, as seen in Fig.~\ref{fig:boostedtopweff} (left).

The performance of the resolved top quark tagger is evaluated using the same methodology as that described in Section~\ref{sec:boostedtopw}. Simulation-to-data correction factors ranging from 1.00 to 1.15 are extracted and  applied to simulated events to account for  differences with data as a function of $\pt$.  Simulated signal events generated in the CMS fast simulation package are corrected in a similar way for differences in tagging performance relative to the full \GEANTfour-based simulations.

\subsection{Identification of initial-state radiation}
\label{sec:isrjets}
In models with $\dm < m_{\PW}$, the LSP is much heavier than the other decay products, and the event has relatively low \met. However, in cases where the $\stopq\stopqbar$~pair recoils against high-\pt~initial-state radiation (ISR), the massive LSP can be either moderately or highly boosted, and there can be relatively large values of \met. To take advantage of this possibility, we try to identify an ISR jet candidate in the event. To this end, we use the set of large-$R$ jets described in Section~\ref{sec:boostedtopw}. The use of such jets improves ISR jet identification by capturing ISR gluon jets that may have undergone splitting to two or more jets that are distributed over a relatively large solid angle. For events having such jets, the large-$R$ jet with the largest value of $\pt \geq 200\GeV$ that fails the ``loose'' working point of the \bq~tagging algorithm (characterized by a tagging efficiency of ${\sim}80\%$, and a misidentification rate of ${\sim}10\%$ for light quarks and gluons, and ${\sim}40\%$ for charm quarks) is tagged as an ISR jet candidate. This ISR jet is then used in SR that are orthogonal to those that require top quark or \PW~boson candidates.

\subsection{Identification of \texorpdfstring{low-\pt \PQb}{low-pt b} quarks}
\label{sec:softbtag}
As previously noted, signal models with small \dm~produce a large fraction of \bq~quarks below the jet \pt~threshold that subsequently fail to be included in the primary jet collection. Identifying these soft quarks can potentially improve our ability to separate signal events from SM background. To this end, we identify soft \bq~hadrons, not associated to jets, by means of a secondary vertex (SV) reconstructed by the inclusive vertex finding algorithm~\cite{Khachatryan:2011wq}. Additional requirements for SV observables are used to suppress background from light-flavor hadrons and jets. These include the distance in the transverse plane between the SV and PV; the significance of this distance; its pointing angle, defined through the scalar product between the distance vector and the $\vec{p}_{\text{SV}}$ direction as $\cos(\overrightarrow{(\text{PV}, \text{SV})},\vec{p}_{\text{SV}})$, where $\vec{p}_{\text{SV}}$ is the total momentum of the tracks associated with the SV; and the number of tracks associated with the SV. The transverse momenta of the tracks associated with an SV are required to sum to $\pt < 20\GeV$, and be separated from any jets (including \bq-tagged jets) by $R > 0.4$. This definition leads to ${\sim}20\%$ efficiency to identify a \bq~hadron in the \pt~range from 10 to 20\GeV, for a misidentification rate less than one percent.
The soft \bq~tagging efficiency in data is measured in a sample dominated by \ttbar~events having an $\Pe\mu$ pair, $\met \geq 50\GeV$, a \bq-tagged jet, and no additional jets. The presence of an additional soft ($\pt < 20\GeV$), nonisolated $\mu$ is used to estimate the fraction of soft \bq~quarks in data. The soft b tagging performance in simulation agrees with the performance in data within 16\%. Simulated signal events produced in the CMS fast simulation package are corrected for differences in soft \bq~tagging relative to \GEANTfour-based simulations.

\section{Search strategy}
\label{sec:srstrategy}
With the final-state signatures of the signals in mind, we select events collected with a \met~trigger and require $\met \geq 250\GeV$ offline.
The SM backgrounds with intrinsic \met~generated through the leptonic decay of a \PW~boson
 are significantly suppressed by rejecting events containing isolated electrons or muons with $\pt \geq 5\GeV$, $\abs{\eta} \leq 2.4$, and $I_\text{rel} \leq 0.1$, or $I_\text{rel} \leq 0.2$, respectively. The contribution from events in which a \PW~boson decays to a $\Pgt$ lepton is suppressed by rejecting events containing isolated $\tauh$ candidates.

In our ``search sample'', defined by the above requirements, the dominant sources of SM background
with intrinsic \met~are \ttbar, \wjets, and \zjets, single top quark, and \ttZ~processes. The contribution from \ttbar and \wjets processes arises from events in which \PW~bosons decay leptonically to produce \met~associated with an energetic neutrino, but the charged lepton either falls outside of the kinematic acceptance, or, even more likely, may be misidentified as a jet after failing to be identified as a lepton. This background is collectively referred to as ``lost lepton" background. Contributions arising from \ttW~and single top quark processes also enter this category at lower levels. The contributions from \zjets and \ttZ~events arise when the \cPZ~boson decays to neutrinos, thereby producing significant \met. Contributions from the QCD multijet process enter the search sample in cases where severe mismeasurements of jet momenta (i.e., jets passing through dead regions, cracks, or transition regions of the detector) produce significant artificial \met, or when neutrinos arise from leptonic decays of heavy-flavor hadrons produced during jet fragmentation.

We define a total of 104 non-overlapping SR with two sets of disjoint baseline selection criteria that are designed specifically for application in the high and low \dm~signals. Tables~\ref{tab:searchregions-hm} and~\ref{tab:searchregions-lm} summarize these criteria for the 51 high \dm~SR and 53 low \dm~SR, respectively.

\subsection{Strategy for high \texorpdfstring{\dm}{Delta m} models}
Based on the final-state signatures of models with $\dm > m_{\text{W}}$, we define a high \dm~baseline selection that requires at least five jets in our primary jet collection ($\nj \geq 5$), of which at least one is \bq-tagged ($\nb \geq 1$). Severely mismeasured high-\pt~jets in multijet events can lead to large values of \met but generally have \ptvecmiss~aligned with one of the higher-\pt~jets in the event. We therefore add the requirement of separation in azimuthal angle between \ptvecmiss~and each of the four jets with largest \pt, $\dphijonetwothreefour \geq 0.5$, which greatly reduces the contribution from this background. Events passing the high \dm~baseline selection are then divided into multiple non-overlapping SR, optimized for the kinematic properties of moderate to high \dm~signal topologies.

In \ljets~\ttbar~events, where most of the \met~is due to the leptonic decay of a single \PW~boson, the transverse mass distribution of the neutrino and \bq~quark from the same top quark decay has an endpoint at the mass of the top quark. To take advantage of this fact, we separate events based on the value of the smallest \bq~quark transverse mass in the event, $\mtb$ (see (2)). In case there are more than two \bq-tagged jets, only the two jets with the highest \bq~tagging discriminant value are considered.
The two resulting sets of events are the \ttbar-depleted high-$\mtb$ category (with $\mtb \geq 175$\GeV), and the \ttbar-enhanced low-$\mtb$ category (with $\mtb < 175$\GeV).

To target signals with moderate values of \dm~that populate the low-$\mtb$ category, we require the presence of at least one resolved top quark and $\nj \geq 7$. The latter condition assures that a signal event would contain at least one radiated jet, providing a boost to the system and thereby increasing the \met~for better discrimination from backgrounds. The high-$\mtb$ category is subdivided into two categories: events that do not contain any top quark or \PW~boson candidates with the requirement $\nj \geq 7$, and events that do contain top quark or \PW~boson candidates, as expected for models with larger values of \dm~and highly boosted top quarks or \PW~bosons. In the latter case, we retain the baseline requirement of $\nj \geq 5$ and define separate SR according to the numbers of candidate merged top quarks ($\nt$), merged \PW~bosons ($\nw$), and resolved top quarks ($\nrt$). All these regions are further subdivided into SR according to the number of \bq-tagged jets, $\nb=1$ or ${\geq}2$, and different ranges of \met.

\subsection{Strategy for low \texorpdfstring{\dm}{Delta m} models}
The low \dm~baseline selection is most appropriate for models with $\dm < m_{\PW}$.
To this end, we select events that have at least two jets, no top quark or \PW~boson candidates, and small $\mtb$ (${<}175\GeV$) when there are \bq-tagged jets present. In addition, we require an ISR jet with $\ptisr \geq 300$\GeV, $ \abs{\eta} \leq 2.4$, and $|\Delta\phi(j_{\text{ISR}},\ptvecmiss)| \geq 2$, where the last requirement suppresses the QCD multijet process. As discussed in Section~\ref{sec:isrjets}, the requirement of an ISR jet provides sensitivity to low \dm~signal topologies, in which intrinsic \met~is generated by the decay of $\stopq\stopqbar$ pairs recoiling against ISR. To further suppress the QCD multijet process, we require $\abs{\Delta\phi(\mathrm{j}_1,\ptvecmiss)} \geq 0.5$,  $\abs{\Delta\phi(\mathrm{j}_{2,3},\ptvecmiss)} \geq 0.15$, where $\text{j}_1, \text{j}_2, \text{j}_3$ are the three leading-\pt~jets. In addition, a measure of significance in \met, defined as $\met / \sqrt{\htt} \equiv\metsig \geq 10~\sqrt{\GeVns{}}$, is required to ensure that \met can only arise from undetectable particles or very rare, extreme mismeasurements.

Events satisfying the above requirements are further subdivided into SR defined by $\nb$, the number of identified secondary vertices $\nsv$, $\ptisr$, and \met. Events with $\nb=0$, a category used for very soft decay products, are further subdivided by ranges of $\nj$, 2 to 5 or ${\geq}6$, $\nsv$, and $\met$, after requiring very high $\ptisr$ to assure a substantial boost to final-state jets which, in turn, enhances the effectiveness of soft \bq~tagging by producing more significantly displaced \bq~hadron decays. The SR with $\nb=\nsv=0$ provide sensitivity to the T2cc model. They may also provide sensitivity to similar final states involving lighter quarks but we have not studied these cases. Events with $\nb \geq 1$ are subdivided according to the scalar sum of the \pt~of the leading and subleading (if one is present) \bq-tagged jets, $\ptbonetwo$, to take advantage of the softer \bq~jet \pt spectrum expected from the low \dm~models relative to the SM background.

\begin{table}[!ht]
\centering
\topcaption{\label{tab:searchregions-hm} Summary of the 51 non-overlapping search regions that mainly target high \dm~signal. The high \dm~baseline selection is $\nj \geq 5$, $\met \geq 250\GeV$, no leptons, $\nb\geq1$, and $\dphijonetwothreefour\geq0.5$.}
\begin{tabular}{*{6}{c}}
	\hline
	\multicolumn{6}{c}{$\mtb < 175$\GeV}                                                                                   \\
	\hline
	$\nj$                    & $\nt$   & $\nw$   & $\nrt$  & $\nb$      & $\met$\,[\GeVns{}]                                      \\
	\hline
	${\geq}7$                  & ${\geq}0$ & ${\geq}0$ & ${\geq}1$ & 1, ${\geq}2$ & 250-300, 300-400, 400-500, ${\geq}500$            \\
	\hline	\\[-1.2ex]

	\multicolumn{6}{c}{$\mtb \geq 175$\GeV}                                                                                   \\
	\hline
	$\nj$                    & $\nt$   & $\nw$   & $\nrt$  & $\nb$      & $\met$\,[\GeVns{}]                                      \\
	\hline
	${\geq}7$                  & 0       & 0       & 0       & 1, ${\geq}2$ & 250-350, 350-450, 450-550, ${\geq}550$            \\
	\hline
	\multirow{4}{*}{${\geq}5$} 	& ${\geq}1$       & 0       		& 0       			& \multirow{4}{*}{1} & 550-650, ${\geq}650$                                    \\
								& 0       		& 0       		& ${\geq}1$	  		&              		& 250-350, 350-450, 450-550, 550-650, ${\geq}650$ \\
								& ${\geq}1$ 		& ${\geq}1$  		& 0       			& 					& ${\geq}550$                                             \\
								& 0       		& ${\geq}1$ 		& ${\geq}1$			& 					& 250-350, 350-450, 450-550, ${\geq}550$                       \\
	\hline
	\multirow{9}{*}{${\geq}5$}	& 1       		& 0       		& 0       			& \multirow{9}{*}{${\geq}2$}   	& 550-650, ${\geq}650$                                    \\
								& 0       		& 1       		& 0       			&     				& 250-350, 350-450, 450-550, 550-650, $\geq 650$ \\
								& 0       		& 0       		& 1       			&  					& 250-350, 350-450, 450-550, 550-650, ${\geq}650$ \\
								& 1				& 1   	    	& 0       			&  					& ${\geq}550$                                             \\
								& 0       		& 1       		& 1       		   	&   				& 250-350, 350-450, 450-550, ${\geq}550$                       \\
								& 1       		& 0       		& 1       		   	&        			& 250-350, 350-450, ${\geq}450$                                  \\
								& ${\geq}2$ 		& 0             & 0                	&        			& ${\geq}250$                                             \\
								& 0       		& ${\geq}2$ 		& 0                	&        			& ${\geq}250$                                             \\
								& 0       		& 0       	 	& ${\geq}2$    		&        			& ${\geq}250$                                            \\
	\hline
\end{tabular}
\end{table}

\begin{table}[!ht]
\centering
\topcaption{\label{tab:searchregions-lm} Summary of the 53 non-overlapping search regions that mainly target low \dm~signal. The low \dm~baseline selection is $\nj \geq 2$, $\met \geq 250\GeV$, no leptons, $\nt=\nw=\nrt=0$, $\mtb < 175\GeV$ (when applicable), $\abs{\Delta\phi(\mathrm{j}_1,\ptvecmiss)} \geq 0.5, ~~ \abs{\Delta\phi(\mathrm{j}_{2,3},\ptvecmiss)} \geq 0.15$, and an ISR jet with $\ptisr \geq 300$\GeV, $ \abs{\eta} \leq 2.4$, $\abs{\Delta\phi(j_{\text{ISR}},\ptvecmiss)} \geq 2$, and $\metsig \geq 10\,\sqrt{\GeVns{}}$.}
\begin{tabular}{*{6}{c}}
	\hline
	$\nj$              & $\nb$                    & $\nsv$             & $\ptisr$\,[\GeVns{}]         & $\ptbonetwo$\,[\GeVns{}]      & $\met$\,[\GeVns{}]                           \\
	\hline
	2--5              & \multirow{4}{*}{0}       & 0                  & \multirow{4}{*}{${\geq}500$} & \multirow{4}{*}{---} & 450--550, 550--650, 650--750, ${\geq}750$ \\
	${\geq}6$            &                          & 0                  &                         &                    & 450--550, 550--650, 650--750, ${\geq}750$ \\
	2--5              &                          & ${\geq}1$            &                         &                    & 450--550, 550--650, 650--750, ${\geq}750$ \\
	${\geq}6$            &                          & ${\geq}1$            &                         &                    & 450--550, 550--650, 650--750, ${\geq}750$ \\
	\hline
	\multirow{5}{*}{${\geq}2$}   & \multirow{5}{*}{1}  & 0                  & 300--500               & 20--40            & 300--400, 400--500, 500--600, ${\geq}600$ \\
	&                          & 0                  & 300--500               & 40--70            & 300--400, 400--500, 500--600, ${\geq}600$ \\
	&                          & 0                  & ${\geq}500$                  & 20--40            & 450--550, 550--650, 650--750, ${\geq}750$ \\
	&                          & 0                  & ${\geq}500$                  & 40--70            & 450--550, 550--650, 650--750, ${\geq}750$ \\
	&                          & ${\geq}1$            & ${\geq}300$                  & 20--40            & 300--400, 400--500, ${\geq}500$            \\
	\hline
	${\geq}2$            & \multirow{6}{*}{${\geq}2$} & \multirow{6}{*}{${\geq}0$} & 300--500               & 40--80            & 300--400, 400--500, ${\geq}500$            \\
	${\geq}2$            &                          &                          & 300--500               & 80--140           & 300--400, 400--500, ${\geq}500$            \\
	${\geq}7$            &                          &                          & 300--500               & ${\geq}140$             & 300--400, 400--500, ${\geq}500$            \\
	${\geq}2$            &                          &                          & ${\geq}500$                  & 40--80            & 450--550, 550--650, ${\geq}650$            \\
	${\geq}2$            &                          &                          & ${\geq}500$                  & 80--140           & 450--550, 550--650, ${\geq}650$            \\
	${\geq}7$            &                          &                          & ${\geq}300$                  & ${\geq}140$             & 450--550, 550--650, ${\geq}650$            \\
	\hline
\end{tabular}
\end{table}

\section{Background estimation}
The contribution of each SM background process to the search sample is estimated through measurements of dedicated control data events that are translated to predictions for event counts in the corresponding SR with the aid of simulation. The strategy makes use of methods described in Ref.~\cite{Sirunyan:2016jpr}.

\subsection{Estimation of the lost-lepton background}
The lost-lepton (LL) background is estimated from a single-lepton control sample that is based on a sample of events collected with the same \met~trigger as the search sample. We create a relatively pure single lepton sample (``$1\ell$'') by inverting the electron or muon veto requirements described in Section~\ref{sec:srstrategy}. More than 90\% of the events in these samples contain a single lepton, while the remainder contain two or more leptons. Studies of simulated events indicate that event kinematic variables for different lepton flavors are sufficiently similar to provide a collective estimate of LL backgrounds from a single control sample. Potential contamination by signal is suppressed by requiring $m_{\mathrm{T}}(\vec{p}_{\mathrm{T}}(\ell),\ptvecmiss) < 100$\GeV, consistent with the expectation for a \PW~boson decay. In events with more than one identified lepton, the one used in this calculation is selected randomly. The selection criteria applied to the single-lepton control sample are the same as those used in the search sample, with the exception of top quark and \PW~boson multiplicity, as discussed below.

The LL estimation in each SR is based upon the event count in corresponding single-lepton control regions (CR). The count is translated to a prediction in the SR by means of a transfer factor obtained from simulation, as follows:
\begin{equation}
N_\text{pred}^{\text{LL}} = \tfll \, N_{\text{data}}(1\ell) ,
\end{equation}
where $N_{\text{data}}(1\ell)$ corresponds to the event count observed in the relevant single-lepton CR in data, and the transfer factor, $\tfll$, translates $N_{\text{data}}(1\ell)$ to a background prediction in the SR, $N_\text{pred}^{\text{LL}}$, and is defined as:
\begin{equation}
\tfll = \frac{N_{\text{MC}}(0\ell)}{N_{\text{MC}}(1\ell)} ,
\label{eqn:tfll}
\end{equation}
where $N_{\text{MC}}(0\ell)$ and $N_{\text{MC}}(1\ell)$ are the LL yields found for simulated events in the search and single-lepton samples, respectively, that include contributions from \ttbar and \PW+jets events, as well as smaller contributions from single top quark and \ttW~processes.

To improve the statistical uncertainty of this background estimation, CR relevant to the high \dm~SR are combined for all top quark and \PW~boson multiplicities in both data and simulation.
 The top quark and \PW~boson tagger results for the simulated events are corrected by the simulation-to-data correction factors discussed in Section~\ref{sec:reco}. Simulation is used to extrapolate these results to each SR with its particular top quark and \PW~boson multiplicity. The selection efficiency for each of the other search variables is estimated directly from data in the single-lepton sample.

\subsection{Estimation of the \texorpdfstring{$\znunu$}{Znunu} background}
An important source of background in the search arises from events in which a \Z~boson, produced in association with jets, decays to neutrinos that carry away large \met. Two methods are traditionally used \cite{Khachatryan:2016xvy,Khachatryan:2016oia} to estimate this background. The first method uses an event sample dominated by \zll$+$jets events, in which the \Z~bosons have kinematic properties very similar to those in the search sample, after correcting for the difference in acceptance between charged lepton pairs and pairs of neutrinos. One drawback in this is that these events are statistically limited, especially in the stringently defined SR often used in SUSY searches. To overcome this limitation, the second method utilizes \gjets~events, in which the \gjets~process has similar LO Feynman diagrams to the $\Z+$jets process, but is more copious than the \zll$+$jets by about a factor of five. To use this sample requires taking into account the differences in quark--boson couplings and the fact that the \Z~boson is very massive. Fortunately, these differences are substantially reduced for the high-$\pt$ bosons in this search.

Considering the pros and cons of the two methods led us to use a hybrid method to estimate the \znunu~background that makes use of both procedures. We use the \zll$+$jets sample to get the normalization of the \znunu+jets~background. This is done in different ranges of $\nb$~and $\nsv$~to account for dependence on heavy-flavor production. Meanwhile, the \gjets~events are used to correct the \met~distributions of simulated events. The \zll~sample is collected with dielectron and dimuon triggers that require the leading electron (muon) to have $\pt \geq 25\,~(20)$\GeV, and the subleading electron (muon) to have $\pt \geq 15\,(10)$\GeV. The leptons must have $\abs{\eta} \leq 2.4$ to be within the acceptance of the tracker. The \gjets events are collected with a single-photon trigger and an offline selection of $\pt \geq 200$\GeV and $\abs{\eta} \leq 2.5$ for the leading photon. To suppress potential contributions from signals and to avoid overlap with the search sample we consider only the events with $\met < 200\GeV$. The transverse momentum of the boson, as determined from the lepton pair or the photon, is added vectorially to \ptvecmiss to emulate the kinematic properties of the \znunu+jets~process. The modified \met, denoted by \metll and \metg for the \zll+jets~and \gjets processes, respectively, is used to calculate the relevant kinematic variables.

The prediction for the \znunu~background in any particular SR is given by:
\begin{equation}
N^{\text{pred}}_{\znunuM} = N^{\text{sim}}_{\znunuM}~R_{\cPZ}~S_{\gamma},
\end{equation}
where $N^{\text{sim}}_{\znunuM}$ is the expected number of \znunu~events in simulation, $R_{\cPZ}$ is the flavor-dependent \zjets~normalization factor measured using the \zll~events, and $S_{\gamma}$ is the correction factor for the \met~distribution as extracted from the \gjets events in data.

The factor $R_{\cPZ}$ is calculated by comparing the observed and expected \zll~yields after applying the baseline selection criteria, with the exception of the requirements on the azimuthal angles between jets and \met. The latter are omitted to retain more events and hence reduce the statistical uncertainty in the $R_{\cPZ}$ estimation, after first confirming that this omission does not bias the result. To increase the purity of the \zll~sample, we require the dilepton invariant mass to lie within the \cPZ~boson mass window of $80 \leq M_{\ell\ell} < 100$\GeV. To probe similar phase space as in the search sample, the $\pt$ of the dilepton system is required to be above 200\GeV. The normalization of the nonnegligible \ttbar~contamination is estimated from the sidebands of the \cPZ~boson mass window of $50 \leq M_{\ell\ell} < 80$ and $M_{\ell\ell} \geq 100$\GeV. Small contributions from $\cmsSymbolFace{t}\cPZ$, \ttbarZ, \wz, and \zz~production, estimated from simulation, are included in the \zll~sample when measuring $R_{\cPZ}$; whereas contributions from $\cmsSymbolFace{t}\PW$, \ttbarW, and \ww~are included in the simulated sample used to obtain the normalization factor for the \ttbar contamination. To account for effects related to heavy-flavor production, $R_{\cPZ}$ is calculated separately for the $\nb$ and $\nsv$ requirements used in different SR. The $R_{\cPZ}$ values are consistent with unity. The uncertainty in $R_{\cPZ}$, ranging from 1 to 29\%, comes mainly from the event counts in data and simulation after implementing the selections, and is treated as a systematic uncertainty in the prediction of the \znunu~background.

The correction factor $S_{\gamma}$ is calculated in each of the search categories via a comparison of the \metg distributions of \gjets events in simulation and data. The event count from simulation is first normalized to the number of events in data after applying the appropriate \dm~baseline selections. The $S_{\gamma}$ factor is estimated separately for each SR, to account for any potential mismodeling of the search variables in simulation. As for the LL background estimation, good agreement between simulation and data for the performance of the top quark and \PW~boson taggers provides a way for us to combine CR for all multiplicities of top quarks and \PW~bosons to calculate $S_{\gamma}$, thereby improving the statistical uncertainty of the result. We then use simulation to extrapolate these results to each SR with its particular top quark and \PW~boson multiplicity, after correcting events using the simulation-to-data correction factors discussed in Section~\ref{sec:reco}.

An underlying assumption of the hybrid estimation is that any differences between data and simulation in the \met~distributions for $\znunu$ events should be compatible with those in the \metg~distributions for photon events. We checked this assumption by comparing the ratios of data to simulation for the \metll and \metg distributions of \zll+jets and \gjets samples, respectively. Residual differences in data and simulation can arise in the process of object reconstruction or as a result of the absence of higher-order corrections in simulation. Observed differences are included in the systematic uncertainties of the \znunu~prediction.

\subsection{Estimation of the QCD multijet background}
The background originating from the QCD multijet process generally constitutes less than 10\% of the total SM background in the SR. It is estimated using a control region in data, consisting of events collected with the same trigger as that used in the search. A sample dominated by the QCD multijet process is then obtained by requiring the azimuthal angle between any one of the three leading jets and \met, $\dphijonetwothree$, to be smaller than 0.1.

We again translate the observation in the control sample to a prediction in the search sample by means of transfer factors obtained from simulation. Each transfer factor is defined by the ratio between the number of simulated QCD multijet events satisfying the SR selection on the azimuthal angles of the four leading jets and \met, to the number of simulated QCD multijet events satisfying $\dphijonetwothree  \leq  0.1$. Contributions from other SM processes to the QCD multijet control sample are subtracted after normalizing the simulation to data in dedicated control samples. The estimation is made in each SR. To improve the statistical uncertainty of the prediction, we combine the CR over $\nt$, $\nw$, and $\nrt$, in data and in simulation. In similarity with the estimations of the LL and \znunu~ backgrounds, we extrapolate in top quark and \PW~boson multiplicity using simulation that is corrected for differences in the top quark and \PW~boson tagging performance with respect to data. In the low \dm~SR categories, we also combine regions of \met~in the QCD multijet control sample when yields are limited for the CR defined by $\nb \geq1$, and we assign an uncertainty for the combination based on the data-to-simulation ratios observed in CR with $\nb=0$.

The dominant source of events originating from QCD multijet processes that populate the SR is from the severe mismeasurement of the \pt~of one or more jets in the event, which translates to large values of artificial \met. The level of mismeasurement can be parameterized via the response variable $\rjet$, defined as the ratio of the reconstructed \pt~of the most mismeasured jet in the event to its generated \pt, computed without including the loss of visible momentum due to neutrinos. The most mismeasured jet is selected based on the jet with greatest absolute difference between the reconstructed and generated $\pt$. In data, we construct the observable $\prjet$, defined as the ratio of the $\pt$ of a given jet to the magnitude of the vector sum of $\ptvec$ and the total $\ptvecmiss$ of the event, which offers a measure of the true jet response. The jet closest in $\phi$ to $\ptvecmiss$ is chosen for this calculation. Mismeasurement correction factors are extracted by comparing the $\prjet$ distributions in data and simulation. The correction factors are parameterized as functions of $\rjet$ and flavor of the most mismeasured jet. The corrections range from 4 to 82\%, and are applied to the simulation on an event-by-event basis.

Due to the large production cross section of the QCD multijet process, samples of simulated QCD multijet events entering the stringently defined SR have limited statistics. To increase it, we use a method that lets each event from the original sample appear multiple times. To this end, we use event ``smearing'', whereby a ``new" event is created by randomly assigning $\rjet$ values to the leading two jets, ranked by their generated jet \pt, and then recalculating all search variables based on the ``smeared'' jets. The $\rjet$ values are sampled from inclusive $\rjet$ distributions binned in both generated jet \pt and jet flavor in a region centered on the original $\rjet$ value. Each original event is smeared 100 times, and the statistical uncertainty in evaluated quantities is estimated through a bootstrapping procedure \cite{tEFR82a} that utilizes 50 pseudo-experiments. We assign a systematic uncertainty of 50\% based on the measured difference in the distribution of the azimuthal angles between the leading jets and \met before and after smearing. This accounts for any potential bias introduced in this method.

\subsection{Estimation of ``rare'' SM backgrounds}
Contributions from diboson (\ww, \wz, and \zz) processes are relatively small compared to the other backgrounds discussed above, and mainly affect the SR in the low \dm~analysis. We therefore estimate this background directly from simulation, with an uncertainty in the production cross section of 50\% \cite{Khachatryan:2015sha,CMS:2016kxu,CMS:2015vwc}. The \ttZ~contribution is also generally very small due to the rarity of this process. However, in SR requiring more than one top quark or \PW~boson, this process can constitute a significant fraction of the total background due to the strong suppression of all other SM backgrounds. The \ttZ~simulation is validated using a three-lepton control sample, obtained using single-lepton triggers, requiring the presence of exactly three leptons (electrons or muons) that satisfy $\pt \geq 40\GeV$ for the leading lepton, $\pt \geq 20\GeV$ for the second and third lepton, and no additional lepton with $\pt \geq 10\GeV$. We further require at least five jets, of which at least two are b-tagged. The same-flavor, opposite-sign lepton pair with the highest dilepton $\pt$ is assumed to originate from \Z~boson decay. We require the presence of such a pair with the invariant mass near the \Z~boson mass (80--100\GeV) and $\pt$ greater than 100\GeV to probe boson kinematic properties similar to those in the search sample. The region outside the Z boson mass window is used to constrain the \ttbar~background. We find that yields in simulated \ttZ~agree with those observed in data. An uncertainty of 24\% is assigned to the normalization of the \ttZ~background in the SR, based on the statistical uncertainty of the simulation-to-data correction factor obtained from this comparison. To assess any potential bias related to the extrapolation from the \Z boson \pt ($\pt(\Z)$) range of the control sample to that of the search sample, we evaluate the \ttZ~simulation-to-data correction factors with different requirements on the reconstructed $\pt(\Z)$, and find the \pt-binned correction factors to be consistent with the inclusive correction factor evaluated for $\pt  \geq  100$\GeV. Theoretical uncertainties related to the choice of PDF and renormalization ($\mu_{\text{R}}$) and factorization ($\mu_{\text{F}}$) scales are found to be up to 28\% in simulated events. 

\subsection{Validation of the background methods in data}
\begin{figure}[ht]
\centering
\includegraphics[width=0.48\textwidth]{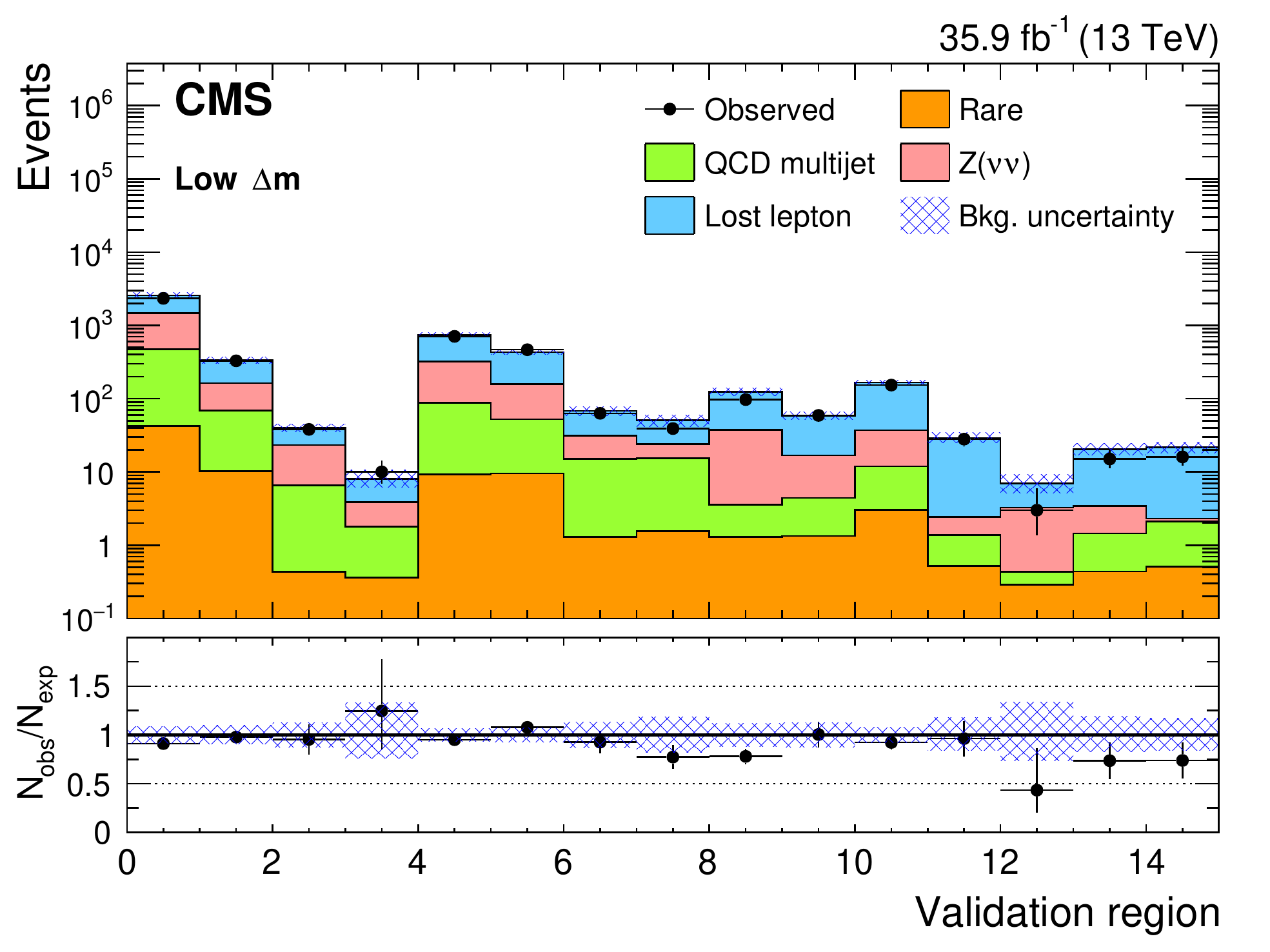}
\includegraphics[width=0.48\textwidth]{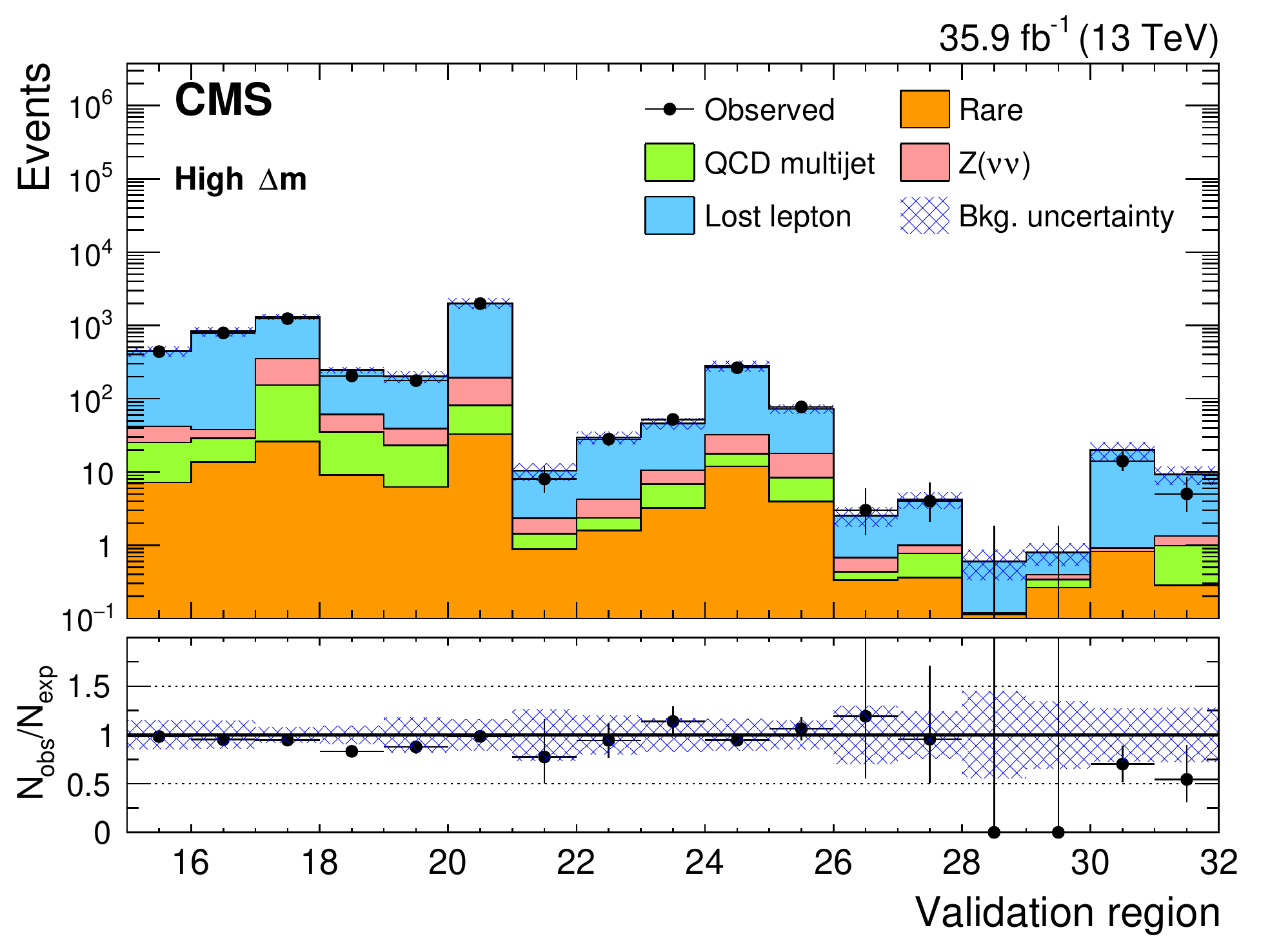}
\caption{Observed data and SM background predictions in the low-\met~validation sample for the low \dm~(left) and high \dm~(right) selections. Ratios of the observed to SM predicted event counts derived from control regions are shown in the lower panel of each plot. The shaded blue band represents the statistical uncertainty combined with the systematic uncertainty resulting from the top quark and \PW~boson tagging correction factors on the background prediction.}
\label{fig:bkgvalidation}
\end{figure}

The background estimation strategy is validated in a data control sample that is non-overlapping to the samples used in the search and in the background estimation described above. The validation sample uses the same selection as the search sample, but focuses on low-\met~regions that are not utilized in the search. The requirement in high \dm~event categories of $\mtb \geq 175\GeV$ is also inverted when selecting events with at least two top quarks or \PW~bosons to increase the statistical power of the validation exercise. Potential signal contamination in the validation regions is negligible. Figure~\ref{fig:bkgvalidation} shows the predicted backgrounds and the observed data in the validation regions. The selections defining each bin are summarized in Table \ref{tab:validation_regions}. The SM prediction is consistent with the observed data, and no indication of a bias is found in the background estimation strategy.

\begin{table}[!!htbp]
\centering
\topcaption{\label{tab:validation_regions} Summary of the validation region selections. The top part of the table (rows 0-14) corresponds to regions for the low \dm~selections, whereas the bottom part (rows 15-31) corresponds to regions for the high \dm~selections.}
\resizebox*{\textwidth}{!}{
\begin{tabular}{llc}
	\hline
	Region & Selection & \met\,[\GeVns{}] \\
	\hline
	0 & $\nb=0$, $\nsv=0$, $\ptisr \geq 500$\GeV, $2\leq\nj\leq5$  & 250--400\\
	1 & $\nb=0$, $\nsv=0$, $\ptisr \geq 500$\GeV, $\nj\geq6$    & 250--400\\
	2 & $\nb=0$, $\nsv\geq1$, $\ptisr \geq 500$\GeV, $2\leq\nj\leq5$ & 250--400\\
	3 & $\nb=0$, $\nsv\geq1$, $\ptisr \geq 500$\GeV, $\nj\geq6$   & 250--400\\
	4 & $\nb=1$, $\nsv=0$, $300 \leq \ptisr < 500$\GeV, $\ptb < 40$\GeV & 250--300\\
	5 & $\nb=1$, $\nsv=0$, $300 \leq \ptisr < 500$\GeV, $40 \leq \ptb < 70$\GeV & 250--300\\
	6 & $\nb=1$, $\nsv=0$, $\ptisr \geq 500$\GeV, $\ptb < 40$\GeV  & 250--400\\
	7 & $\nb=1$, $\nsv=0$, $\ptisr \geq 500$\GeV, $40 \leq \ptb < 70$\GeV & 250--400\\
	8 & $\nb=1$, $\nsv\geq1$, $\ptb < 40$\GeV                  & 250--300\\
	9 & $\nb\geq2$, $300 \leq \ptisr < 500$\GeV, $\ptbonetwo < 80$\GeV & 250--300\\
	10 & $\nb\geq2$, $300 \leq \ptisr < 500$\GeV, $80 \leq \ptbonetwo < 140$\GeV & 250--300\\
	11 & $\nb\geq2$, $300 \leq \ptisr < 500$\GeV, $\ptbonetwo \geq 140$\GeV, $\nj\geq7$ & 250--300\\
	12 & $\nb\geq2$, $\ptisr \geq 500$\GeV, $\ptbonetwo < 80$\GeV & 250--400\\
	13 & $\nb\geq2$, $\ptisr \geq 500$\GeV, $80 \leq \ptbonetwo < 140$\GeV & 250--400\\
	14 & $\nb\geq2$, $\ptisr \geq 500$\GeV, $\ptbonetwo \geq 140$\GeV, $\nj\geq7$ & 250--400\\
	\hline
	15 & $\nb=1$, $\mtb < 175$\GeV, $\nj\geq7$, $\nrt\geq1$    & 200--250\\
	16 & $\nb\geq2$, $\mtb < 175$\GeV, $\nj\geq7$, $\nrt\geq1$ & 200--250\\
	17 & $\nb=1$, $\nj\geq7$, $\nt=0$, $\nrt=0$, $\nw=0$      & 200--250\\
	18 & $\nb\geq2$, $\nj\geq7$, $\nt=0$, $\nrt=0$, $\nw=0$   & 200--250\\
	19 & $\nb=1$, $\nt\geq1$, $\nrt=0$, $\nw=0$ & 200--450\\
	20 & $\nb=1$, $\nt=0$, $\nrt\geq1$, $\nw=0$ & 200--250\\
	21 & $\nb=1$, $\nt\geq1$, $\nrt=0$, $\nw\geq1$ & 200--450\\
	22 & $\nb=1$, $\nt=0$, $\nrt\geq1$, $\nw\geq1$ & 200--250\\
	23 & $\nb\geq2$, $\nt=1$, $\nrt=0$, $\nw=0$ & 200--450\\
	24 & $\nb\geq2$, $\nt=0$, $\nrt=1$, $\nw=0$ & 200--250\\
	25 & $\nb\geq2$, $\nt=0$, $\nrt=0$, $\nw=1$ & 200--250\\
	26 & $\nb\geq2$, $\nt=1$, $\nrt=0$, $\nw=1$ & 200--450\\
	27 & $\nb\geq2$, $\nt=0$, $\nrt=1$, $\nw=1$ & 200--250\\
	28 & $\nb\geq2$, $\nt=1$, $\nrt=1$, $\nw=0$ & 200--250\\
	29 & $\nb\geq2$, $\nt\geq2$, $\nrt=0$, $\nw=0$, $\mtb < 175$\GeV or $\met < 250$\GeV & ${\geq}200$\\
	30 & $\nb\geq2$, $\nt=0$, $\nrt\geq2$, $\nw=0$, $\mtb < 175$\GeV & 200--250\\
	31 & $\nb\geq2$, $\nt=0$, $\nrt=0$, $\nw\geq2$, $\mtb < 175$\GeV & 200--250\\
	\hline
\end{tabular}
}
\end{table}

\section{Systematic uncertainties}
As described in the preceding section, our strategy for estimating the background relies on translating event counts from data control regions to search regions by means of transfer factors obtained from simulation. These transfer factors, as well as the signal predictions, are therefore sensitive to a variety of systematic uncertainties related to the modeling of the experimental apparatus, particle kinematic properties, and  theoretical models. We discuss the relevant sources of uncertainty below, and summarize their effects on the predictions in the SR in Table~\ref{tab:sysunc}.

\begin{itemize}
\item Choosing the size of control samples in data and simulation can lead to statistical limitations.
\item Important systematic effects can arise from the dependence of the top quark and \PW~boson tagging performance on the modeling of the \ttbar~topology and showering of bottom quarks and partons from decays of \PW~bosons to quarks. We assess a systematic uncertainty in the modeling of the \ttbar~topology by comparing the tagging efficiencies in simulation between \ttbar~samples generated using \POWHEG~and \MADGRAPH. The relative effect on the tagging efficiencies ranges from 1 to 4\%. Uncertainties related to the choice of scheme in parton showering are evaluated by comparing the tagging and mistagging efficiencies in simulation between the \PYTHIA and \HERWIGpp~shower models. Differences of 5 to 25\% and 5 to 40\% are seen for tagging and misidentification, respectively.
We also evaluate the impact of heavy flavor jet multiplicity on the data control sample used to define the misidentification correction factors by comparing the selection for $\nb=0$ to the nominal selection of $\nb \geq1$. The observed difference of 20\% is assigned as a systematic uncertainty.

The statistical precision in the tagging correction factors of the top quarks and \PW\ bosons ranges from 1 to 25\%. The definition of the generator-level matching scheme used when applying the tagging correction factors, as well as top \pt~reweighting, are found to have very small impact on the top quark and \PW\ boson tagging performance. The sources of uncertainty discussed in this paragraph are collectively referred to as the ``remaining sources'' in Table~\ref{tab:sysunc}.

\item In simulating rare background processes and signal, a 16\% uncertainty is assessed to account for simulation-to-data differences in soft \bq~tagging performance based on comparisons in control regions.

\item Correction factors applied to simulation to account for differences in lepton selection efficiency between data and simulation have associated systematic uncertainties related to changes in the performance of the tracker over the data-taking period.

\item Changes in $\mu_{\text{R}}$ and $\mu_{\text{F}}$ scales, PDF, and the strong coupling strength, $\alpha_{S}$ impact rare background predictions more significantly than background estimation from control samples in data that often benefit from partial or full cancellation of these uncertainties. The effect of unknown higher-order effects  on any  predicted event count is estimated by varying simultaneously $\mu_{\text{R}}$ and $\mu_{\text{F}}$ by a factor of two, as detailed in Refs.~\cite{Catani2003zt,Cacciari2003fi}, and ranges from 1 to 10\%. The uncertainty related to the choice of $\text{PDF}$, including the uncertainty in $\alpha_{S}$, is obtained as the standard deviation in 100 variations of the NNPDF3.0~\cite{Ball:2014uwa} PDF, and ranges from 1 to 28\%. The PDF systematics are evaluated only for background processes.

\item A 2.5\% uncertainty is assigned to the integrated luminosity measured by the CMS experiment for the 2016 data-taking period~\cite{CMS-PAS-LUM-17-001}, and affects the simulation-based prediction of rare SM background and signal processes.

\item A variety of other sources of systematic uncertainties include  the corrections for \bq~tagging performance, jet energy scale and resolution, which also affect the \met~in the event, and reweighting of events for pileup. These sources are generally of much smaller importance compared to the other sources.

The magnitude of the systematic uncertainties is typically small, except for the most stringently defined SR affected by larger statistical uncertainties in the corresponding data control samples. The background estimation strategy, which translates event counts from data control samples with kinematic properties very similar to the corresponding SR, benefits from partial or full cancellation of many of the above sources of systematic uncertainty.

\end{itemize}

\begin{table}[htbp]
\topcaption{\label{tab:sysunc} Range of systematic uncertainties [\%] in the prediction across the different search regions. ``Rare" column includes diboson and \ttZ~processes. ``Signal" column shows the range of systematic uncertainties representative of the full set of models shown in~Fig.~\ref{fig:diagram}.
}
\centering
\begin{tabular}{l*{5}{c}}
\hline
Source & Lost lepton & \znunu & QCD multijet & Rare & Signal \\
\hline
$\Pe/\mu$ veto             & 1--5  & 2--3 & 1--6 & 1--8 & 1--5 \\
$\tauh$ veto     & 1--7   & 2--3 & 1--7 & 1--7 & 1--8 \\
\bq~tagging: heavy flavor  & 1--4 & 1--5 & 1--14 & 1--6 & 1--10 \\
\bq~tagging: light flavor  & 1--7 & 1--15 & 1--16 & 1--8 & 1--20 \\
Soft \bq~tagging           & --- & --- & --- & 1--16 & 1--16 \\
Jet energy scale           & 1--30 & 1--25 & 1--6 & 1--31 & 1--35 \\
\met~resolution            & 1--13& 1--18& 1--5& 1--30 & 1--48 \\
\ttbar~normalization       & 1--8 & --- & --- & --- & --- \\
\PW+jets normalization      & 1--11 & --- & --- & --- & --- \\
Top quark \pt              & 1--24 & --- & --- & --- & --- \\
Sample size (MC)           & 1--100 & 1--100 & 2--100 & 4--100 & 2--100 \\
Sample size (data CR)      & 2--100 & 1--100 & 3--100 & --- & --- \\
$R_{\cPZ}$                 & --- & 1--29 & --- & --- & --- \\
$\cPZ/\gamma$ difference   & --- & 1--23 & --- & --- & --- \\
Background subtraction     & --- & --- & 13--71 & --- & --- \\
Jet response tail          & --- & --- & 1--14 & --- & --- \\
\met integration           & --- & --- & 1--51 & --- & --- \\
Smearing closure           & --- & --- & 50 & --- & --- \\
Pileup reweighting         & 1--12 & 1--12 & 1--12 & 1--16 & 1--15 \\
Integrated luminosity      & --- & --- & --- & 2.5 & 2.5 \\
Cross section              & --- & --- & --- & 24--50 & --- \\
PDF and $\alpha_S$ dependence    & 1--10 & --- & --- & 1--10 & --- \\
$\mu_{\text{R}}/\mu_{\text{F}}$ dependence
                           & 1--7 & --- & --- & 1--28 & 1--9 \\
Merged tagging             &  &  &  &  &  \\
\qquad  Generator                & $<$3 & --- & --- & $<$4 & $<$4 \\
\qquad  Parton showering         & 1--22 & 1--23 & 1--42 & 1--25 & 1--34 \\
\qquad  Sample size (data)       & 1--5 & 1--7 & 1--11 & 1--3  & 1--3 \\
\qquad  Mistag $\nb$               & 1--22 & 1--18 & 1--18 & 1--6 & 1--5 \\
\qquad  Remaining sources        & 1--24 & --- & --- & 1--27 & 1--32 \\
Resolved tagging                 & & & & & \\
\qquad  Generator                & $<$1 & --- & --- & $<$1 & $<$3 \\
\qquad  Parton showering         & 1--12 & --- & --- & 1--16 & 1--31 \\
\qquad  Remaining sources        & 1--18 & 1--17 & 1--17 & 1--16 & 1--20 \\
\hline
\end{tabular}
\end{table}

\section{Results and interpretation}
\begin{figure}[!!pt]
	\centering
		\includegraphics[width=0.7\textwidth]{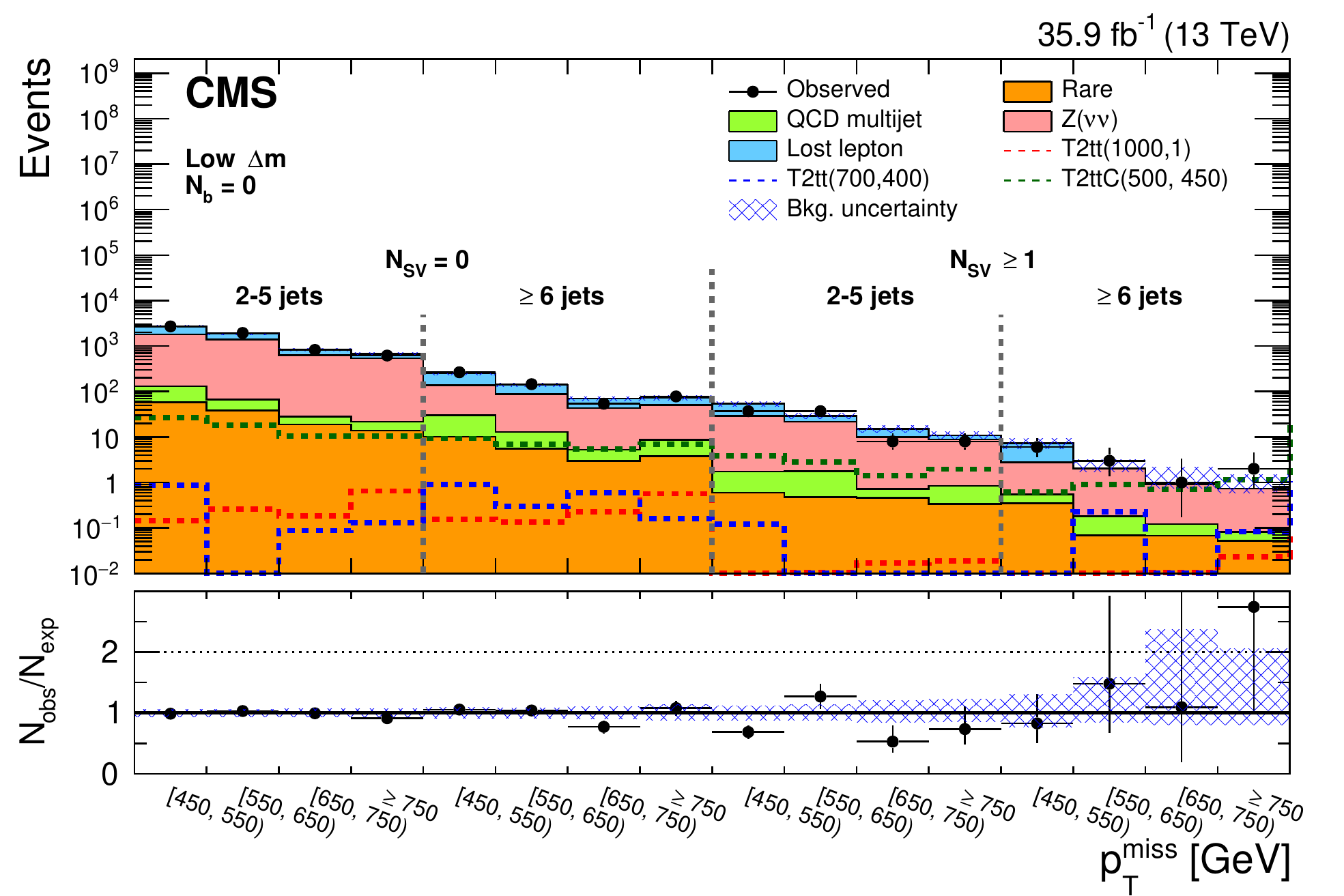}\\
	\caption{\label{fig:results-lm-1} Observed data events and SM background predictions for the low \dm~search regions with $\nb=0$. Ratios of the observed to SM predicted event counts are shown in the lower panel of each plot. The shaded blue band represents the combined statistical and systematic uncertainties on the SM predictions.  Units are\GeV.}
\end{figure}

\begin{figure}[!!thp]
	\centering
		\includegraphics[width=0.7\textwidth]{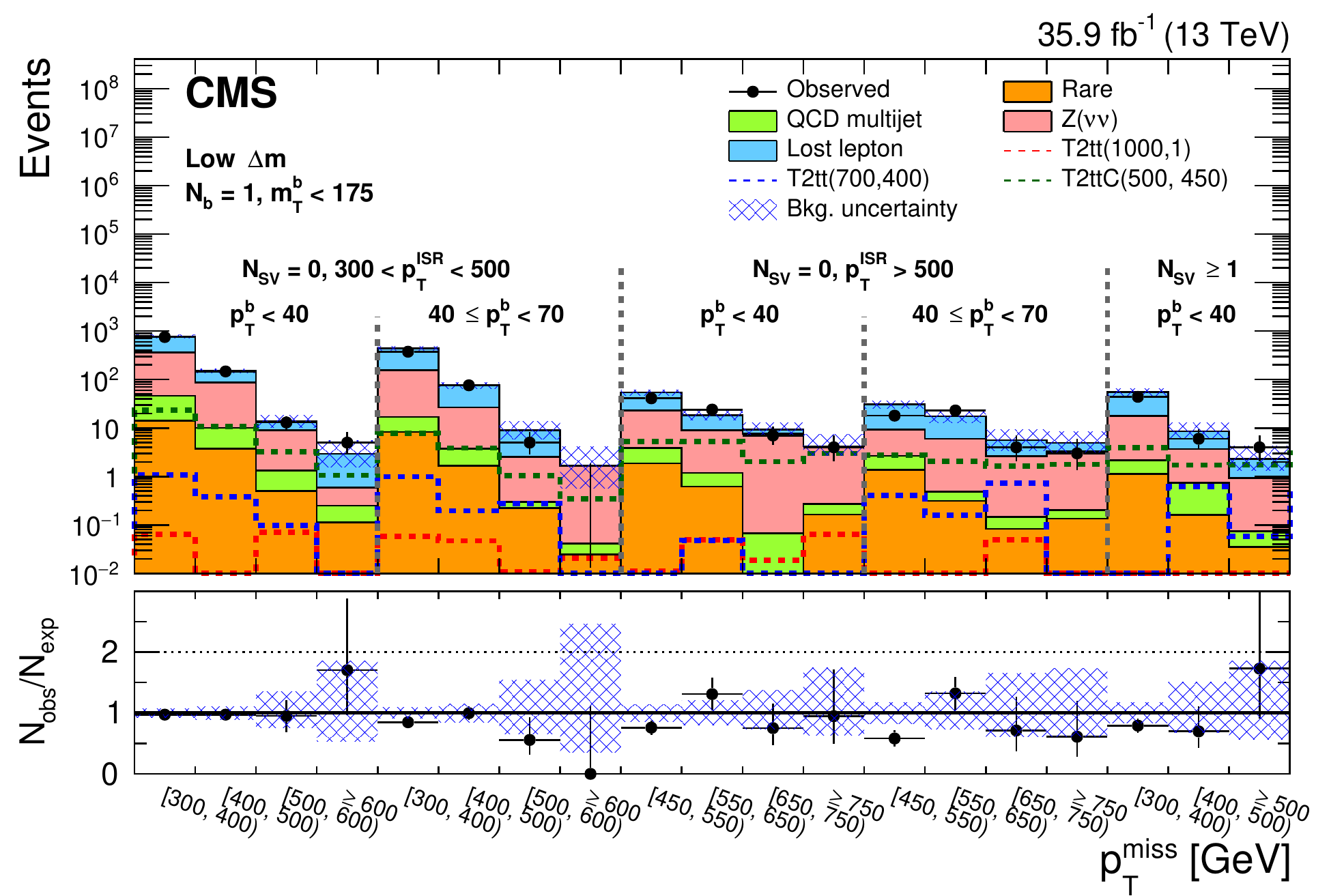}\\
		\includegraphics[width=0.7\textwidth]{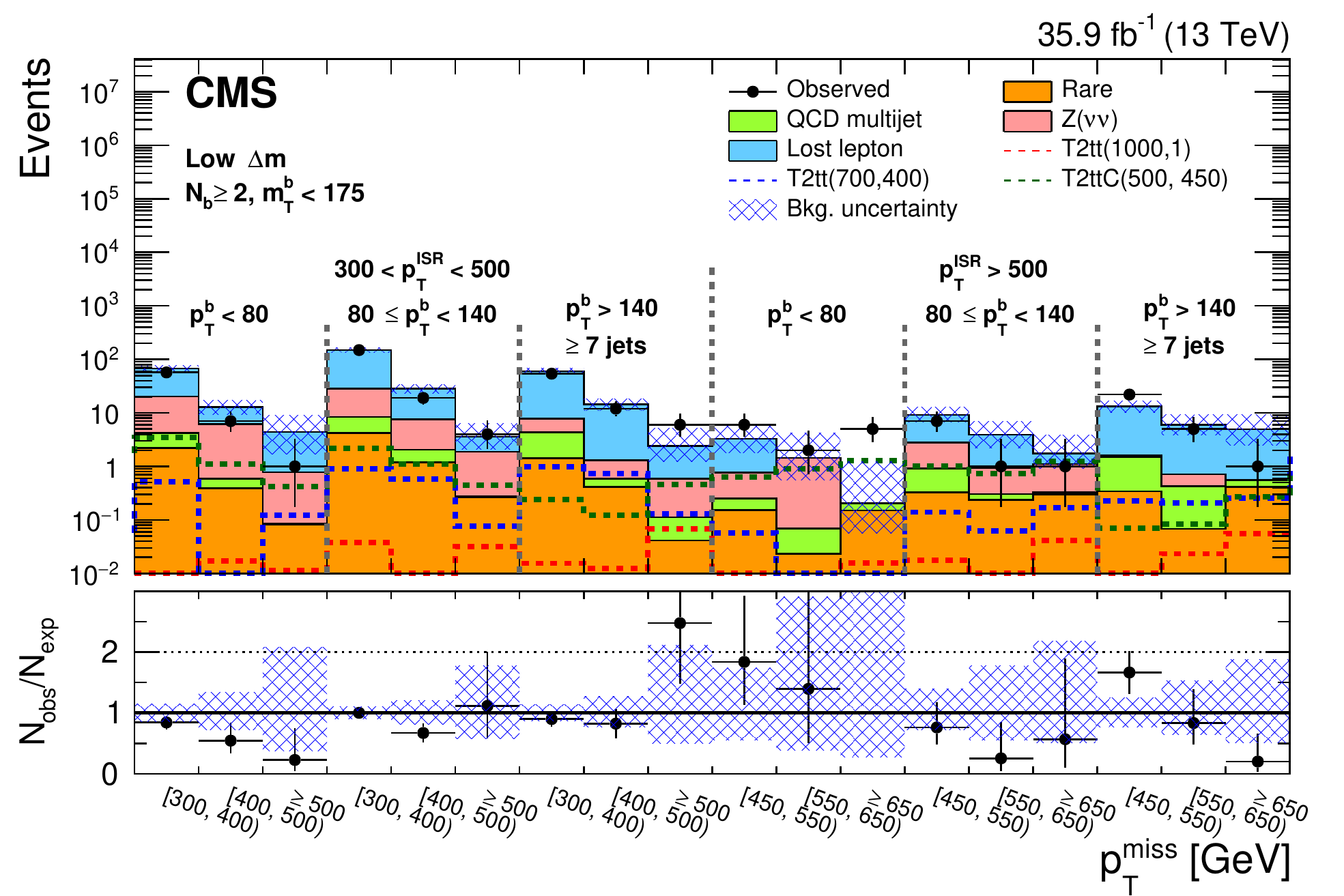}	
	\caption{\label{fig:results-lm-2} Observed data events and SM background predictions for the low \dm~search regions with $\nb=1$ (upper), and $\nb\geq2$ (lower). Ratios of the observed to SM predicted event counts are shown in the lower panel of each plot. The shaded blue band represents the combined statistical and systematic uncertainties on the SM predictions. The one SR for which the result is not visible in the lower panel is discussed in the text. Units are\GeV.}
\end{figure}

\begin{figure}[!!p]
	\centering
		\includegraphics[width=0.7\textwidth]{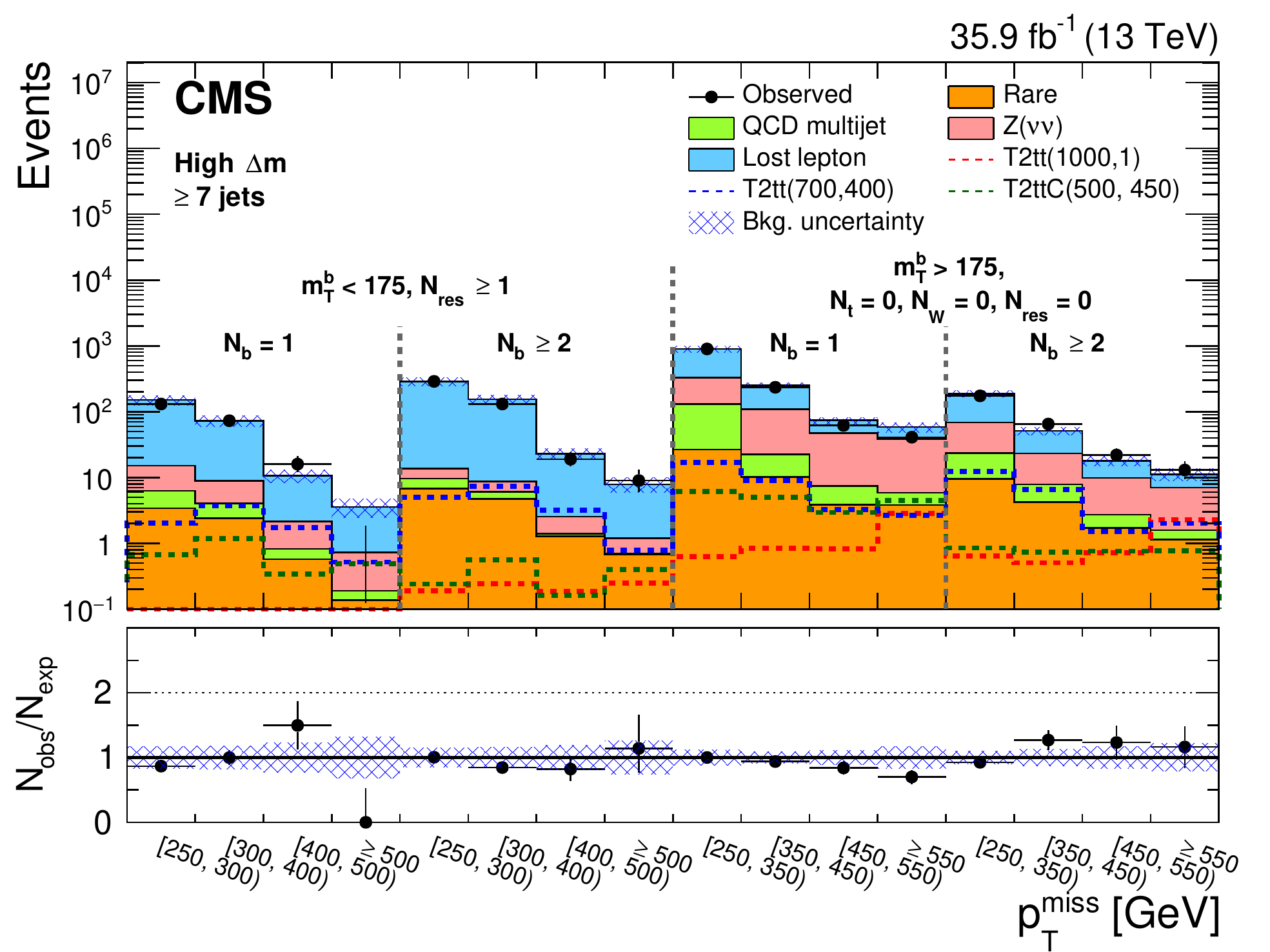} 
                \includegraphics[width=0.7\textwidth]{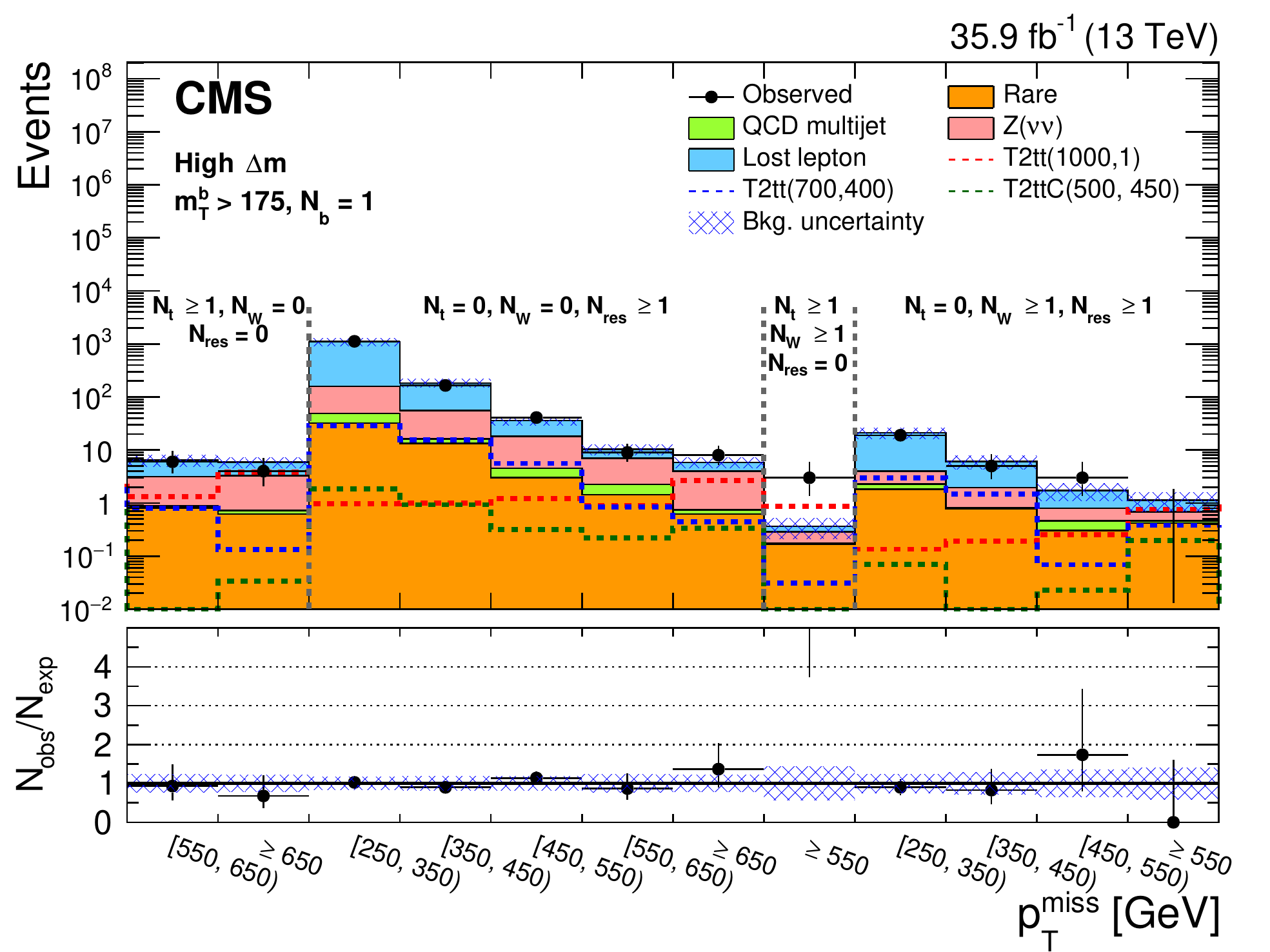} 

	\caption{\label{fig:results-hm-1} Observed data events and SM background predictions for the high \dm~search regions with $\nj \geq 7$ (upper), and $\mtb>175~\GeV$ and $\nb=1$ (lower). Details of the selection applied is displayed on each plot.
 Ratios of the observed to SM predicted event counts are shown in the lower panel of each plot. The shaded blue band represents the statistical and systematic uncertainty on the prediction. The one SR for which the result is not visible in the lower panel is discussed in the text. Units are\GeV.}
\end{figure}

\begin{figure}[!!tp]
	\centering
		\includegraphics[width=0.7\textwidth]{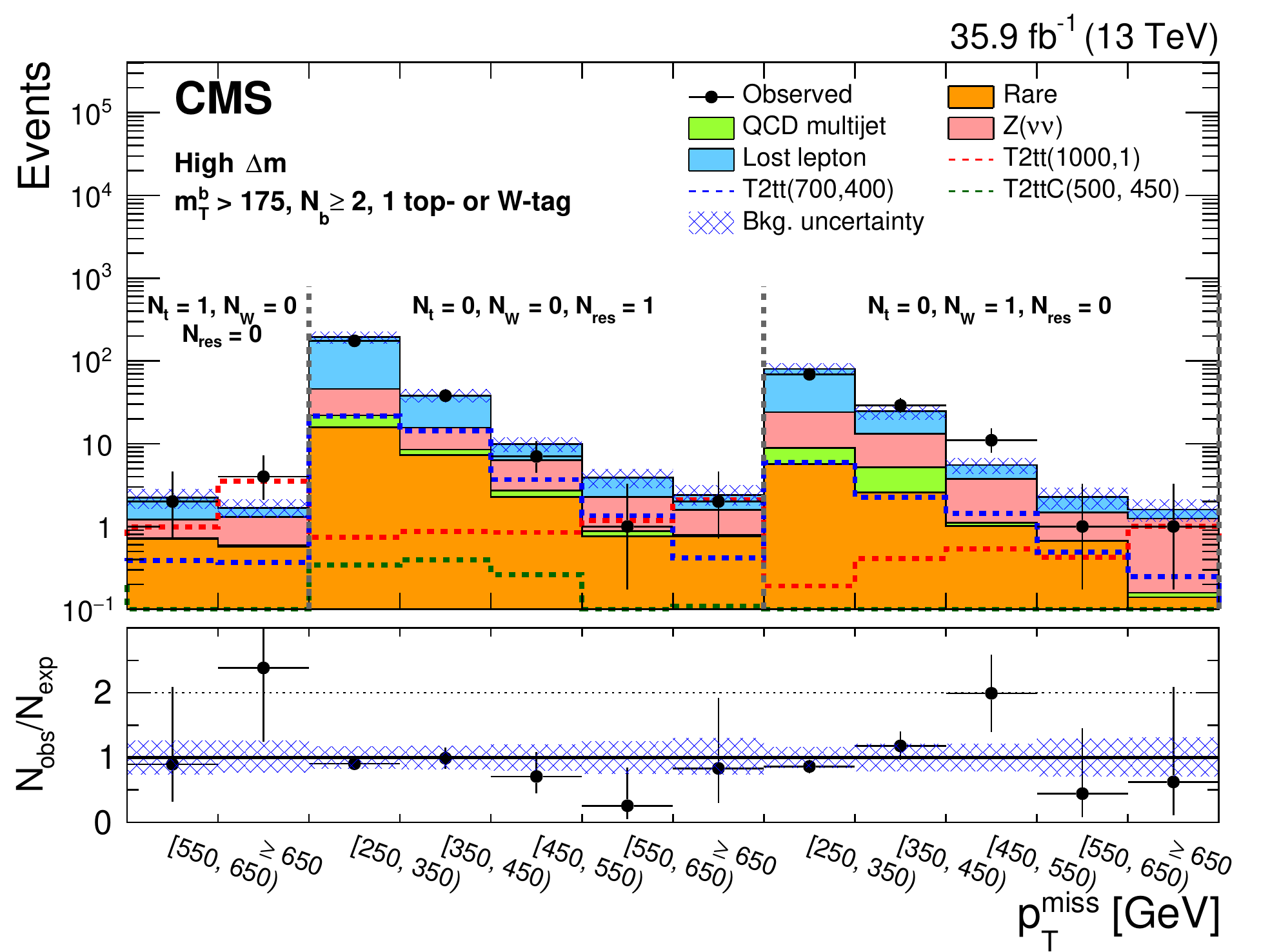}
		\includegraphics[width=0.7\textwidth]{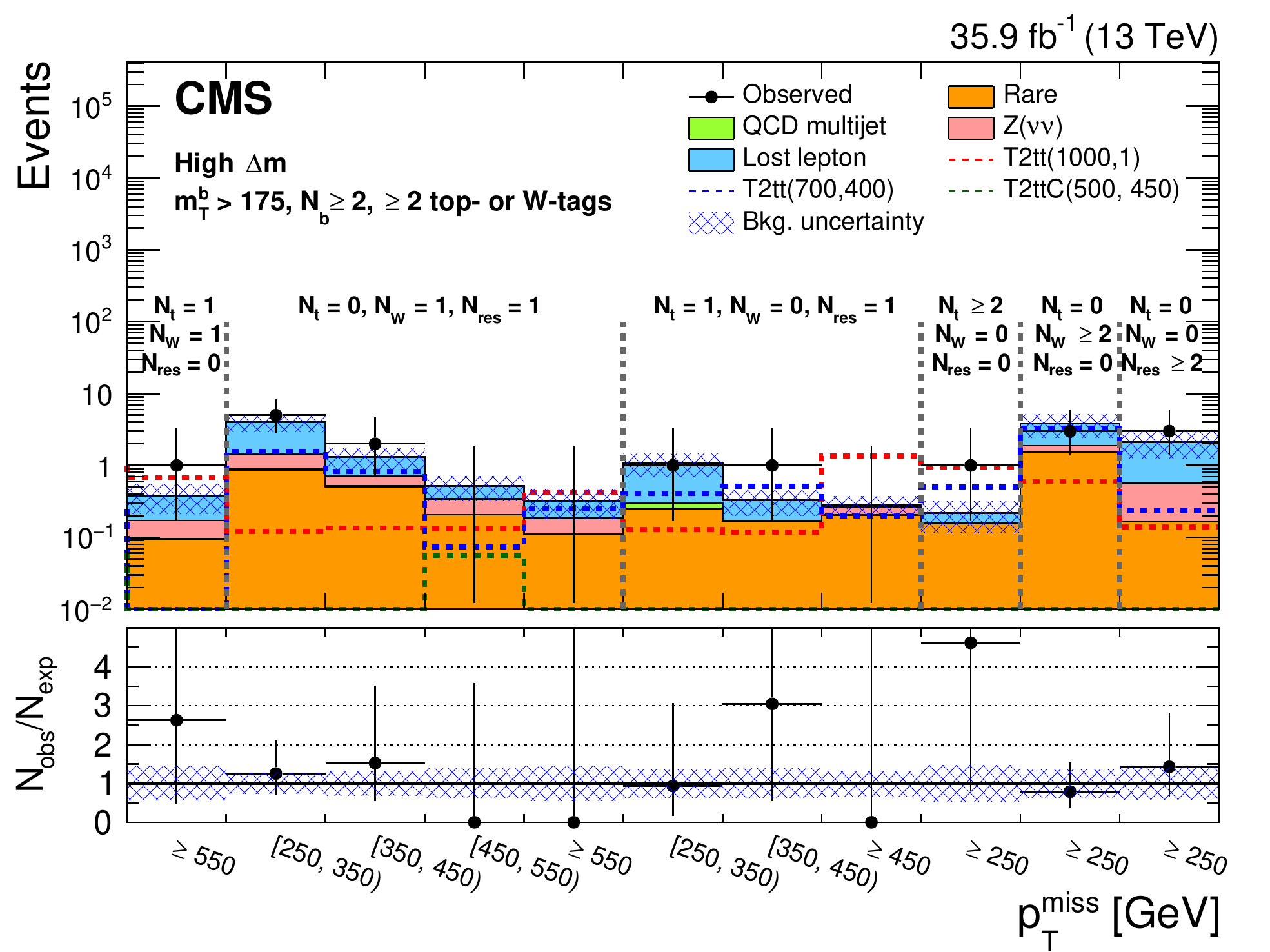}	
	\caption{\label{fig:results-hm-2} Observed data events and SM background predictions for the high \dm~search regions with one (upper), and at least two (lower) top quark or W boson candidates. Details of the selection applied is displayed on each plot.
 Ratios of the observed to SM predicted event counts are shown in the lower panel of each plot. The shaded blue band represents the statistical and systematic uncertainty on the prediction.  Units are\GeV.}
\end{figure}

The event counts observed in data and those predicted for SM backgrounds are summarized graphically in Figs.~\ref{fig:results-lm-1}-\ref{fig:results-hm-2}, and numerically in Tables~\ref{tab:data-vs-pred-lm} and \ref{tab:data-vs-pred-hm}.
 The observed event counts are in general agreement with the predictions. The two search regions with most significant discrepancies are the low \dm~SR defined by the selection $\nb\geq2$, $\mtb<175$\GeV, $\ptisr \geq 500$\GeV, $\ptbonetwo < 80$\GeV, $\met \geq 650$\GeV, and the high \dm~SR defined by $\nb=1$, $\mtb \geq 175$\GeV, $\nt\geq1$, $\nrt=0$, $\nw\geq1$, $\met \geq 550$\GeV. For these two SR, observed excesses over the predicted event counts correspond to local significances of 2.3 and 1.9 standard deviations, respectively. These can be attributed to statistical fluctuations of the SM backgrounds alone, given the number of search regions employed in this analysis.

\begin{table}[!!htbp]
\centering
\topcaption{\label{tab:data-vs-pred-lm} Predicted background yields and the observation in different search regions for the low \dm~analysis. The total uncertainty is given for each background prediction.}
\resizebox*{0.60\textwidth}{!}{
\begin{tabular}{*{7}{c}}
\hline
\met\,[\GeVns{}]  &  Lost lepton  &  \znunu  & Rare &  QCD multijet &  Total SM  &  $N_\text{data}$  \\
\hline\\[-1.2ex]
\multicolumn{7}{c}{$\nb=0$, $\nsv=0$, $\ptisr \geq 500$\GeV, $2\leq\nj\leq5$} \\
\hline
450-550 & $ 935\pm73 $  & $ 1670\pm120 $  & $ 58\pm29 $  & $ 73\pm37 $  & $ 2740\pm180 $  & 2704 \\
550-650 & $ 498\pm39 $  & $ 1318\pm84 $  & $ 38\pm19 $  & $ 28\pm14 $  & $ 1880\pm110 $  & 1942 \\
650-750 & $ 202\pm19 $  & $ 597\pm43 $  & $ 19\pm10 $  & $ 9.6\pm4.9 $  & $ 828\pm55 $  & 823 \\
$\geq 750$ & $ 135\pm14 $  & $ 520\pm38 $  & $ 14\pm7 $  & $ 7.9\pm4.2 $  & $ 676\pm46 $  & 618 \\
\hline\\[-1.2ex]
\multicolumn{7}{c}{$\nb=0$, $\nsv=0$, $\ptisr \geq 500$\GeV, $\nj\geq6$} \\
\hline
450-550 & $ 115\pm12 $  & $ 106\pm10 $  & $ 10\pm5 $  & $ 20\pm10 $  & $ 251\pm22 $  & 265 \\
550-650 & $ 52\pm6 $  & $ 74\pm7 $  & $ 5.5\pm2.8 $  & $ 7.3\pm3.8 $  & $ 139\pm12 $  & 145 \\
650-750 & $ 27\pm4 $  & $ 38\pm5 $  & $ 3.0\pm1.6 $  & $ 2.3\pm1.3 $  & $ 70\pm7 $  & 54 \\
$\geq 750$ & $ 21\pm4 $  & $ 42\pm5 $  & $ 3.8\pm2.0 $  & $ 4.9\,^{+6.3}_{-5.2} $  & $ 72\,^{+10}_{-8} $  & 78 \\
\hline\\[-1.2ex]
\multicolumn{7}{c}{$\nb=0$, $\nsv\geq1$, $\ptisr \geq 500$\GeV, $2\leq\nj\leq5$} \\
\hline
450-550 & $ 25\pm5 $  & $ 27\pm3 $  & $ 0.60\pm0.47 $  & $ 1.2\pm0.7 $  & $ 54\pm6 $  & 37 \\
550-650 & $ 7.6\pm2.5 $  & $ 20\pm2 $  & $ 0.47\pm0.37 $  & $ 1.3\,^{+1.2}_{-0.9} $  & $ 29\pm4 $  & 37 \\
650-750 & $ 5.2\,^{+2.7}_{-1.9} $  & $ 9.2\pm1.1 $  & $ 0.46\pm0.40 $  & $ 0.27\,^{+0.29}_{-0.24} $  & $ 15\,^{+3}_{-2} $  & 8 \\
$\geq 750$ & $ 2.0\,^{+2.0}_{-1.1} $  & $ 8.0\pm1.0 $  & $ 0.34\pm0.26 $  & $ 0.50\,^{+0.40}_{-0.34} $  & $ 11\pm2 $  & 8 \\
\hline\\[-1.2ex]
\multicolumn{7}{c}{$\nb=0$, $\nsv\geq1$, $\ptisr \geq 500$\GeV, $\nj\geq6$} \\
\hline
450-550 & $ 4.5\,^{+2.1}_{-1.6} $  & $ 2.2\pm0.4 $  & $ 0.35\pm0.29 $  & $ 0.19\,^{+0.17}_{-0.13} $  & $ 7.2\,^{+2.2}_{-1.7} $  & 6 \\
550-650 & $<$1.08 & $ 1.8\pm0.3 $  & $ 0.07\pm0.05 $  & $ 0.11\,^{+0.10}_{-0.08} $  & $ 2.0\,^{+1.2}_{-0.3} $  & 3 \\
650-750 & $<$1.22 & $ 0.79\pm0.17 $  & $ 0.07\pm0.05 $  & $ 0.05\,^{+0.05}_{-0.04} $  & $ 0.9\,^{+1.3}_{-0.2} $  & 1 \\
$\geq 750$ & $<$0.74 & $ 0.65\pm0.14 $  & $ 0.05\pm0.05 $  & $ 0.03\,^{+0.03}_{-0.02} $  & $ 0.73\,^{+0.77}_{-0.15} $  & 2 \\
\hline\\[-1.2ex]
\multicolumn{7}{c}{$\nb=1$, $\nsv=0$, $\mtb<175$\GeV, $300 \leq \ptisr<500$\GeV, $\ptb<40$\GeV} \\
\hline
300-400 & $ 410\pm38 $  & $ 318\pm29 $  & $ 14\pm7 $  & $ 32\pm17 $  & $ 774\pm57 $  & 753 \\
400-500 & $ 64\pm11 $  & $ 77\pm10 $  & $ 3.8\pm1.9 $  & $ 6.3\pm3.9 $  & $ 151\pm16 $  & 147 \\
500-600 & $ 4.7\,^{+3.9}_{-2.4} $  & $ 7.6\pm2.2 $  & $ 0.5\pm0.3 $  & $ 0.83\pm0.59 $  & $ 14\,^{+5}_{-3} $  & 13 \\
$\geq 600$ & $ 2.4\,^{+2.1}_{-1.3} $  & $ 0.34\,^{+0.79}_{-0.28} $  & $ 0.11\pm0.07 $  & $ 0.14\pm0.11 $  & $ 2.9\,^{+2.5}_{-1.4} $  & 5 \\
\hline\\[-1.2ex]
\multicolumn{7}{c}{$\nb=1$, $\nsv=0$, $\mtb<175$\GeV, $300 \leq \ptisr<500$\GeV, $40 \leq \ptb<70$\GeV} \\
\hline
300-400 & $ 285\pm33 $  & $ 140\pm15 $  & $ 8.3\pm3.8 $  & $ 8.6\pm4.7 $  & $ 442\pm39 $  & 375 \\
400-500 & $ 50\pm10 $  & $ 23\pm4 $  & $ 1.7\pm0.9 $  & $ 2.1\pm1.5 $  & $ 76\pm11 $  & 76 \\
500-600 & $ 6.4\,^{+4.2}_{-2.9} $  & $ 2.3\,^{+1.5}_{-1.0} $  & $ 0.22\pm0.13 $  & $ 0.08\pm0.06 $  & $ 9.0\,^{+4.8}_{-3.1} $  & 5 \\
$\geq 600$ & $<$0.83 & $ 1.6\,^{+1.9}_{-1.1} $  & $ 0.02\pm0.03 $  & $ 0.02\pm0.02 $  & $ 1.7\,^{+2.4}_{-1.1} $  & 0 \\
\hline\\[-1.2ex]
\multicolumn{7}{c}{$\nb=1$, $\nsv=0$, $\mtb<175$\GeV, $\ptisr \geq 500$\GeV, $\ptb<40$\GeV} \\
\hline
450-550 & $ 31\pm6 $  & $ 19\pm4 $  & $ 1.9\pm1.1 $  & $ 2.0\pm1.2 $  & $ 54\pm8 $  & 41 \\
550-650 & $ 9.3\pm3.0 $  & $ 7.8\pm2.0 $  & $ 0.62\pm0.42 $  & $ 0.57\,^{+0.48}_{-0.40} $  & $ 18\pm4 $  & 24 \\
650-750 & $ 1.7\,^{+2.3}_{-1.1} $  & $ 7.5\pm2.2 $  & $ 0.01\pm0.17 $  & $ 0.06\,^{+0.06}_{-0.05} $  & $ 9.3\,^{+3.5}_{-2.5} $  & 7 \\
$\geq 750$ & $<$1.48 & $ 4.0\,^{+2.1}_{-1.5} $  & $ 0.16\pm0.10 $  & $ 0.11\,^{+0.10}_{-0.08} $  & $ 4.2\,^{+3.2}_{-1.5} $  & 4 \\
\hline\\[-1.2ex]
\multicolumn{7}{c}{$\nb=1$, $\nsv=0$, $\mtb<175$\GeV, $\ptisr \geq 500$\GeV, $40 \leq \ptb<70$\GeV} \\
\hline
450-550 & $ 22\pm5 $  & $ 6.6\pm1.7 $  & $ 1.4\pm0.8 $  & $ 1.3\pm0.8 $  & $ 31\pm5 $  & 18 \\
550-650 & $ 11\,^{+6}_{-4} $  & $ 5.5\pm1.8 $  & $ 0.31\pm0.18 $  & $ 0.17\,^{+0.16}_{-0.12} $  & $ 17\,^{+6}_{-5} $  & 23 \\
650-750 & $ 3.0\,^{+2.6}_{-1.6} $  & $ 2.5\,^{+1.9}_{-1.3} $  & $ 0.08\pm0.09 $  & $ 0.06\,^{+0.10}_{-0.06} $  & $ 5.6\,^{+3.7}_{-2.2} $  & 4 \\
$\geq 750$ & $ 1.7\,^{+2.3}_{-1.1} $  & $ 3.1\,^{+2.1}_{-1.5} $  & $ 0.14\pm0.09 $  & $ 0.07\,^{+0.11}_{-0.06} $  & $ 4.9\,^{+3.6}_{-1.9} $  & 3 \\
\hline\\[-1.2ex]
\multicolumn{7}{c}{$\nb=1$, $\nsv\geq1$, $\mtb<175$\GeV, $\ptb<40$\GeV} \\
\hline
300-400 & $ 38\pm8 $  & $ 16\pm5 $  & $ 1.1\pm0.6 $  & $ 1.0\,^{+1.0}_{-0.8} $  & $ 56\,^{+10}_{-9} $  & 44 \\
400-500 & $ 4.9\,^{+3.8}_{-2.5} $  & $ 2.9\pm1.0 $  & $ 0.16\pm0.13 $  & $ 0.58\,^{+0.97}_{-0.54} $  & $ 8.6\,^{+4.4}_{-2.8} $  & 6 \\
$\geq 500$ & $ 1.4\,^{+1.9}_{-1.0} $  & $ 0.86\pm0.31 $  & $ 0.03\pm0.03 $  & $ 0.04\,^{+0.08}_{-0.04} $  & $ 2.3\,^{+2.0}_{-1.0} $  & 4 \\
\hline\\[-1.2ex]
\multicolumn{7}{c}{$\nb\geq2$, $\mtb<175$\GeV, $300 \leq \ptisr<500$\GeV, $\ptbonetwo<80$\GeV} \\
\hline
300-400 & $ 47\pm8 $  & $ 16\pm5 $  & $ 2.2\pm1.0 $  & $ 2.0\,^{+1.8}_{-1.5} $  & $ 68\,^{+10}_{-9} $  & 57 \\
400-500 & $ 6.7\,^{+3.4}_{-2.6} $  & $ 5.5\pm2.4 $  & $ 0.39\pm0.23 $  & $ 0.19\,^{+0.18}_{-0.16} $  & $ 13\pm4 $  & 7 \\
$\geq 500$ & $ 3.6\,^{+4.3}_{-2.7} $  & $ 0.7\,^{+1.1}_{-0.6} $  & $ 0.08\pm0.05 $  & $<$0.01 & $ 4.4\,^{+4.7}_{-2.7} $  & 1 \\
\hline\\[-1.2ex]
\multicolumn{7}{c}{$\nb\geq2$, $\mtb<175$\GeV, $300 \leq \ptisr<500$\GeV, $80\leq\ptbonetwo<140$\GeV} \\
\hline
300-400 & $ 121\pm13 $  & $ 20\pm5 $  & $ 4.2\pm1.7 $  & $ 4.2\pm2.5 $  & $ 149\pm15 $  & 149 \\
400-500 & $ 21\pm5 $  & $ 5.5\pm2.0 $  & $ 1.2\pm0.6 $  & $ 0.9\,^{+1.6}_{-0.9} $  & $ 28\,^{+6}_{-5} $  & 19 \\
$\geq 500$ & $ 1.7\,^{+1.8}_{-1.0} $  & $ 1.6\,^{+1.6}_{-1.0} $  & $ 0.27\pm0.16 $  & $ 0.01\pm0.01 $  & $ 3.6\,^{+2.8}_{-1.5} $  & 4 \\
\hline\\[-1.2ex]
\multicolumn{7}{c}{$\nb\geq2$, $\mtb<175$\GeV, $300 \leq \ptisr<500$\GeV, $\ptbonetwo>140$\GeV, $\nj\geq7$} \\
\hline
300-400 & $ 52\pm8 $  & $ 3.5\,^{+1.9}_{-1.4} $  & $ 1.4\pm0.6 $  & $ 2.9\pm1.8 $  & $ 60\pm8 $  & 54 \\
400-500 & $ 13\pm3 $  & $ 0.7\,^{+1.0}_{-0.5} $  & $ 0.41\pm0.16 $  & $ 0.18\,^{+0.45}_{-0.18} $  & $ 15\,^{+4}_{-3} $  & 12 \\
$\geq 500$ & $ 1.8\,^{+1.9}_{-1.1} $  & $ 0.5\,^{+1.2}_{-0.4} $  & $ 0.04\pm0.15 $  & $ 0.07\,^{+0.19}_{-0.07} $  & $ 2.4\,^{+2.7}_{-1.2} $  & 6 \\
\hline\\[-1.2ex]
\multicolumn{7}{c}{$\nb\geq2$, $\mtb<175$\GeV, $\ptisr \geq 500$\GeV, $\ptbonetwo<80$\GeV} \\
\hline
450-550 & $ 2.5\,^{+2.2}_{-1.4} $  & $ 0.52\,^{+0.46}_{-0.31} $  & $ 0.15\pm0.08 $  & $ 0.1\,^{+0.13}_{-0.09} $  & $ 3.3\,^{+2.4}_{-1.5} $  & 6 \\
550-650 & $<$1.59 & $ 1.4\,^{+1.5}_{-0.9} $  & $ 0.02\pm0.06 $  & $ 0.05\,^{+0.07}_{-0.04} $  & $ 1.4\,^{+2.7}_{-0.9} $  & 2 \\
$\geq 650$ & $<$0.75 & $<$0.33 & $ 0.15\pm0.14 $  & $ 0.06\,^{+0.09}_{-0.06} $  & $ 0.2\,^{+1.0}_{-0.2} $  & 5 \\
\hline\\[-1.2ex]
\multicolumn{7}{c}{$\nb\geq2$, $\mtb<175$\GeV, $\ptisr \geq 500$\GeV, $80\leq\ptbonetwo<140$\GeV} \\
\hline
450-550 & $ 6.4\,^{+3.0}_{-2.2} $  & $ 1.9\,^{+1.3}_{-0.9} $  & $ 0.33\pm0.22 $  & $ 0.58\,^{+0.57}_{-0.47} $  & $ 9.2\,^{+3.7}_{-2.5} $  & 7 \\
550-650 & $ 3.0\,^{+2.6}_{-1.6} $  & $ 0.63\,^{+0.89}_{-0.44} $  & $ 0.24\pm0.16 $  & $ 0.07\,^{+0.06}_{-0.05} $  & $ 3.9\,^{+3.0}_{-1.7} $  & 1 \\
$\geq 650$ & $ 0.7\,^{+1.6}_{-0.6} $  & $ 0.78\,^{+0.87}_{-0.50} $  & $ 0.30\pm0.23 $  & $ 0.03\,^{+0.03}_{-0.02} $  & $ 1.8\,^{+2.1}_{-0.9} $  & 1 \\
\hline\\[-1.2ex]
\multicolumn{7}{c}{$\nb\geq2$, $\mtb<175$\GeV, $\ptisr \geq 500$\GeV, $\ptbonetwo>140$\GeV, $\nj\geq7$} \\
\hline
450-550 & $ 12\pm3 $  & $ 0.12\,^{+0.34}_{-0.12} $  & $ 0.34\pm0.19 $  & $ 1.1\,^{+0.9}_{-0.8} $  & $ 13\pm3 $  & 22 \\
550-650 & $ 5.3\,^{+2.8}_{-2.1} $  & $ 0.29\,^{+0.71}_{-0.25} $  & $ 0.07\pm0.10 $  & $ 0.36\,^{+0.31}_{-0.25} $  & $ 6.0\,^{+3.2}_{-2.1} $  & 5 \\
$\geq 650$ & $ 4.4\,^{+3.8}_{-2.4} $  & $<$0.85 & $ 0.42\pm0.41 $  & $ 0.14\,^{+0.13}_{-0.1} $  & $ 4.9\,^{+4.3}_{-2.4} $  & 1 \\
\hline
\end{tabular}
}
\end{table}

\begin{table}[!!htbp]
\centering
\topcaption{\label{tab:data-vs-pred-hm} Predicted background yields and the observation in different search regions for the high \dm~analysis. The total uncertainty is given for each background prediction.}
\resizebox*{0.60\textwidth}{!}{
\begin{tabular}{*{7}{c}}
\hline
\met\,[\GeVns{}]  &  Lost lepton  &  \znunu  & Rare &  QCD multijet  &  Total SM  &  $N_\text{data}$  \\
\hline\\[-1.2ex]
\multicolumn{7}{c}{$\nb=1$, $\mtb<175$\GeV, $\nj\geq7$, $\nrt\geq1$} \\
\hline
250-300 & $ 136\pm23 $  & $ 8.9\pm2.7 $  & $ 3.4\pm0.9 $  & $ 2.9\pm1.8 $  & $ 151\pm26 $  & 131 \\
300-400 & $ 64\pm12 $  & $ 4.8\pm1.4 $  & $ 2.4\pm0.7 $  & $ 1.7\pm1.1 $  & $ 73\pm13 $  & 73 \\
400-500 & $ 8.5\pm2.1 $  & $ 1.3\pm0.4 $  & $ 0.57\pm0.22 $  & $ 0.25\pm0.17 $  & $ 11\pm2 $  & 16 \\
$\geq 500$ & $ 2.9\pm1.1 $  & $ 0.54\pm0.23 $  & $ 0.14\pm0.08 $  & $ 0.05\pm0.03 $  & $ 3.6\pm1.1 $  & 0 \\
\hline\\[-1.2ex]
\multicolumn{7}{c}{$\nb\geq2$, $\mtb<175$\GeV, $\nj\geq7$, $\nrt\geq1$} \\
\hline
250-300 & $ 274\pm42 $  & $ 4.1\pm1.3 $  & $ 6.8\pm2.0 $  & $ 2.8\pm1.8 $  & $ 288\pm44 $  & 289 \\
300-400 & $ 146\pm23 $  & $ 2.7\pm0.9 $  & $ 4.8\pm1.3 $  & $ 1.3\pm0.9 $  & $ 155\pm24 $  & 131 \\
400-500 & $ 21\pm4 $  & $ 1.2\pm0.5 $  & $ 1.3\pm0.4 $  & $ 0.12\pm0.09 $  & $ 23\pm4 $  & 19 \\
$\geq 500$ & $ 6.7\pm1.9 $  & $ 0.49\pm0.22 $  & $ 0.67\pm0.28 $  & $ 0.03\pm0.02 $  & $ 7.9\pm2.0 $  & 9 \\
\hline\\[-1.2ex]
\multicolumn{7}{c}{$\nb=1$, $\mtb \geq 175$\GeV, $\nj\geq7$, $\nt=0$, $\nrt=0$, $\nw=0$} \\
\hline
250-350 & $ 568\pm63 $  & $ 200\pm18 $  & $ 27\pm8 $  & $ 104\pm69 $  & $ 900\pm100 $  & 899 \\
350-450 & $ 141\pm17 $  & $ 87\pm8 $  & $ 10\pm3 $  & $ 12\pm7 $  & $ 251\pm23 $  & 235 \\
450-550 & $ 27\pm4 $  & $ 40\pm6 $  & $ 3.9\pm1.3 $  & $ 3.6\pm2.1 $  & $ 74\pm8 $  & 62 \\
$\geq 550$ & $ 20\pm4 $  & $ 33\pm8 $  & $ 3.8\pm1.6 $  & $ 2.0\pm1.1 $  & $ 59\pm10 $  & 41 \\
\hline\\[-1.2ex]
\multicolumn{7}{c}{$\nb\geq2$, $\mtb \geq 175$\GeV, $\nj\geq7$, $\nt=0$, $\nrt=0$, $\nw=0$} \\
\hline
250-350 & $ 120\pm15 $  & $ 45\pm6 $  & $ 9.6\pm2.5 $  & $ 14\pm8 $  & $ 188\pm20 $  & 174 \\
350-450 & $ 28\pm5 $  & $ 15\pm3 $  & $ 4.2\pm1.2 $  & $ 3.7\pm2.1 $  & $ 51\pm7 $  & 65 \\
450-550 & $ 8.0\pm2.1 $  & $ 7.2\pm1.6 $  & $ 1.7\pm0.5 $  & $ 1.0\,^{+0.8}_{-0.7} $  & $ 18\pm3 $  & 22 \\
$\geq 550$ & $ 4.2\pm1.3 $  & $ 5.4\pm1.8 $  & $ 1.1\pm0.4 $  & $ 0.45\,^{+0.47}_{-0.37} $  & $ 11\pm2 $  & 13 \\
\hline\\[-1.2ex]
\multicolumn{7}{c}{$\nb=1$, $\mtb \geq 175$\GeV, $\nt\geq1$, $\nrt=0$, $\nw=0$} \\
\hline
550-650 & $ 3.3\pm1.2 $  & $ 2.3\pm0.7 $  & $ 0.81\pm0.26 $  & $ 0.08\pm0.06 $  & $ 6.4\pm1.5 $  & 6 \\
$\geq 650$ & $ 2.6\pm1.0 $  & $ 2.5\pm0.6 $  & $ 0.62\pm0.20 $  & $ 0.11\pm0.08 $  & $ 5.9\pm1.3 $  & 4 \\
\hline\\[-1.2ex]
\multicolumn{7}{c}{$\nb=1$, $\mtb \geq 175$\GeV, $\nt=0$, $\nrt\geq1$, $\nw=0$} \\
\hline
250-350 & $ 930\pm170 $  & $ 110\pm28 $  & $ 32\pm9 $  & $ 17\pm10 $  & $ 1090\pm180 $  & 1120 \\
350-450 & $ 128\pm28 $  & $ 39\pm9 $  & $ 13\pm4 $  & $ 2.9\pm1.8 $  & $ 183\pm33 $  & 165 \\
450-550 & $ 18\pm4 $  & $ 14\pm3 $  & $ 3.0\pm0.9 $  & $ 1.5\pm1.1 $  & $ 36\pm6 $  & 41 \\
550-650 & $ 3.3\pm1.1 $  & $ 4.8\pm1.5 $  & $ 1.4\pm0.5 $  & $ 0.80\pm0.66 $  & $ 10\pm2 $  & 9 \\
$\geq 650$ & $ 1.9\pm0.6 $  & $ 3.2\pm0.8 $  & $ 0.62\pm0.20 $  & $ 0.13\pm0.11 $  & $ 5.8\pm1.3 $  & 8 \\
\hline\\[-1.2ex]
\multicolumn{7}{c}{$\nb=1$, $\mtb \geq 175$\GeV, $\nt\geq1$, $\nrt=0$, $\nw\geq1$} \\
\hline
$\geq 550$ & $ 0.08\pm0.07 $  & $ 0.11\pm0.08 $  & $ 0.17\pm0.07 $  & $ 0.01\pm0.01 $  & $ 0.37\pm0.16 $  & 3 \\
\hline\\[-1.2ex]
\multicolumn{7}{c}{$\nb=1$, $\mtb \geq 175$\GeV, $\nt=0$, $\nrt\geq1$, $\nw\geq1$} \\
\hline
250-350 & $ 17\pm4 $  & $ 1.7\pm0.6 $  & $ 1.8\pm0.6 $  & $ 0.46\pm0.39 $  & $ 21\pm5 $  & 19 \\
350-450 & $ 4.1\pm1.4 $  & $ 1.1\pm0.5 $  & $ 0.79\pm0.26 $  & $ 0.03\pm0.03 $  & $ 6.0\pm1.7 $  & 5 \\
450-550 & $ 0.92\pm0.46 $  & $ 0.34\pm0.14 $  & $ 0.31\pm0.17 $  & $ 0.16\pm0.18 $  & $ 1.7\pm0.6 $  & 3 \\
$\geq 550$ & $ 0.45\pm0.27 $  & $ 0.22\pm0.11 $  & $ 0.42\pm0.31 $  & $ 0.05\pm0.05 $  & $ 1.1\pm0.5 $  & 0 \\
\hline\\[-1.2ex]
\multicolumn{7}{c}{$\nb\geq2$, $\mtb \geq 175$\GeV, $\nt=1$, $\nrt=0$, $\nw=0$} \\
\hline
550-650 & $ 1.0\pm0.5 $  & $ 0.48\pm0.19 $  & $ 0.7\pm0.2 $  & $ 0.03\pm0.03 $  & $ 2.2\pm0.6 $  & 2 \\
$\geq 650$ & $ 0.38\,^{+0.27}_{-0.22} $  & $ 0.71\pm0.23 $  & $ 0.56\pm0.17 $  & $ 0.03\,^{+0.03}_{-0.02} $  & $ 1.7\pm0.4 $  & 4 \\
\hline\\[-1.2ex]
\multicolumn{7}{c}{$\nb\geq2$, $\mtb \geq 175$\GeV, $\nt=0$, $\nrt=1$, $\nw=0$} \\
\hline
250-350 & $ 148\pm26 $  & $ 24\pm6 $  & $ 16\pm4 $  & $ 6.2\pm4.3 $  & $ 194\pm32 $  & 175 \\
350-450 & $ 23\pm5 $  & $ 7.2\pm1.8 $  & $ 7.3\pm2.0 $  & $ 1.2\pm0.8 $  & $ 38\pm7 $  & 38 \\
450-550 & $ 3.6\pm1.1 $  & $ 3.6\pm1.0 $  & $ 2.3\pm0.6 $  & $ 0.46\pm0.40 $  & $ 9.9\pm2.0 $  & 7 \\
550-650 & $ 1.6\pm0.6 $  & $ 1.4\pm0.5 $  & $ 0.76\pm0.25 $  & $ 0.12\pm0.13 $  & $ 3.9\pm1.0 $  & 1 \\
$\geq 650$ & $ 0.82\,^{+0.45}_{-0.34} $  & $ 0.80\pm0.25 $  & $ 0.75\pm0.35 $  & $ 0.04\,^{+0.05}_{-0.04} $  & $ 2.4\,^{+0.7}_{-0.6} $  & 2 \\
\hline\\[-1.2ex]
\multicolumn{7}{c}{$\nb\geq2$, $\mtb \geq 175$\GeV, $\nt=0$, $\nrt=0$, $\nw=1$} \\
\hline
250-350 & $ 56\pm9 $  & $ 15\pm4 $  & $ 5.7\pm1.7 $  & $ 3.2\pm1.9 $  & $ 80\pm13 $  & 69 \\
350-450 & $ 11\pm2 $  & $ 8.0\pm2.3 $  & $ 2.6\pm0.8 $  & $ 2.6\pm1.8 $  & $ 25\pm5 $  & 29 \\
450-550 & $ 1.8\pm0.6 $  & $ 2.6\pm0.8 $  & $ 1.0\pm0.4 $  & $ 0.10\pm0.09 $  & $ 5.5\pm1.2 $  & 11 \\
550-650 & $ 0.78\pm0.36 $  & $ 0.80\pm0.34 $  & $ 0.67\pm0.39 $  & $<$0.01 & $ 2.3\pm0.7 $  & 1 \\
$\geq 650$ & $ 0.36\,^{+0.25}_{-0.20} $  & $ 1.1\pm0.4 $  & $ 0.14\pm0.09 $  & $ 0.02\pm0.02 $  & $ 1.6\pm0.5 $  & 1 \\
\hline\\[-1.2ex]
\multicolumn{7}{c}{$\nb\geq2$, $\mtb \geq 175$\GeV, $\nt=1$, $\nrt=0$, $\nw=1$} \\
\hline
$\geq 550$ & $ 0.21\pm0.14 $  & $ 0.08\pm0.05 $  & $ 0.10\pm0.03 $  & $<$0.01 & $ 0.38\pm0.17 $  & 1 \\
\hline\\[-1.2ex]
\multicolumn{7}{c}{$\nb\geq2$, $\mtb \geq 175$\GeV, $\nt=0$, $\nrt=1$, $\nw=1$} \\
\hline
250-350 & $ 2.6\pm0.8 $  & $ 0.51\pm0.23 $  & $ 0.86\pm0.28 $  & $ 0.05\pm0.06 $  & $ 4.0\pm1.1 $  & 5 \\
350-450 & $ 0.60\pm0.29 $  & $ 0.20\pm0.11 $  & $ 0.51\pm0.19 $  & $ 0.01\pm0.01 $  & $ 1.3\pm0.4 $  & 2 \\
450-550 & $ 0.17\pm0.13 $  & $ 0.14\pm0.08 $  & $ 0.21\pm0.07 $  & $<$0.01 & $ 0.52\pm0.20 $  & 0 \\
$\geq 550$ & $ 0.14\pm0.11 $  & $ 0.07\pm0.06 $  & $ 0.11\pm0.05 $  & $<$0.01 & $ 0.32\pm0.14 $  & 0 \\
\hline\\[-1.2ex]
\multicolumn{7}{c}{$\nb\geq2$, $\mtb \geq 175$\GeV, $\nt=1$, $\nrt=1$, $\nw=0$} \\
\hline
250-350 & $ 0.77\pm0.33 $  & $<$0.01 & $ 0.25\pm0.15 $  & $ 0.05\pm0.06 $  & $ 1.1\pm0.4 $  & 1 \\
350-450 & $ 0.16\pm0.11 $  & $<$0.01 & $ 0.17\pm0.06 $  & $<$0.01 & $ 0.33\pm0.13 $  & 1 \\
$\geq 450$ & $ 0.01\pm0.01 $  & $ 0.06\pm0.04 $  & $ 0.20\pm0.08 $  & $<$0.01 & $ 0.28\pm0.09 $  & 0 \\
\hline\\[-1.2ex]
\multicolumn{7}{c}{$\nb\geq2$, $\mtb \geq 175$\GeV, $\nt\geq2$, $\nrt=0$, $\nw=0$} \\
\hline
$\geq 250$ & $ 0.06\pm0.06 $  & $<$0.01 & $ 0.16\pm0.07 $  & $<$0.01 & $ 0.22\pm0.10 $  & 1 \\
\hline\\[-1.2ex]
\multicolumn{7}{c}{$\nb\geq2$, $\mtb \geq 175$\GeV, $\nt=0$, $\nrt\geq2$, $\nw=0$} \\
\hline
$\geq 250$ & $ 1.9\pm0.8 $  & $ 0.35\pm0.22 $  & $ 1.5\pm0.7 $  & $<$0.01 & $ 3.8\pm1.4 $  & 3 \\
\hline\\[-1.2ex]
\multicolumn{7}{c}{$\nb\geq2$, $\mtb \geq 175$\GeV, $\nt=0$, $\nrt=0$, $\nw\geq2$} \\
\hline
$\geq 250$ & $ 1.5\pm0.7 $  & $ 0.39\pm0.2 $  & $ 0.17\pm0.13 $  & $<$0.01 & $ 2.1\pm0.9 $  & 3 \\
\hline
\end{tabular}
}
\end{table}

The statistical interpretations of the results in terms of exclusion limits for signal models being considered in this analysis are based on a binned likelihood fit to the data, which takes into account the predicted background and signal yields in the SR. The extraction of exclusion limits is based on a modified frequentist approach using the CL$_\mathrm{s}$ criterion~\cite{Junk:1999kv,Read:2002hq} under the asymptotic approximation for the test statistic~\cite{CLs,Cowan:2010js}. All of the SR, and their corresponding CR, are fitted simultaneously to determine the signal cross section excluded at a 95\% confidence level (CL) for each signal point. Models for signal in which the 95\% CL upper limit on the production cross section falls below the theoretical value (based on NLO+NLL calculations) are considered excluded by this analysis.

The inclusion of the single-lepton CR in the likelihood fit ensures that any signal contamination in the CR is taken into account, through estimates of the corresponding signal, by the fit. The systematic uncertainties assigned to the signal and background are treated as nuisance parameters in the fit. The term ``nuisance parameter'' refers to a variable of little physical interest which however needs to be taken into account in order to have precise modeling of parameters that are of physical interest. Statistical uncertainties due to the limited number of simulated events are uncorrelated among all regions and backgrounds. The statistical uncertainties in background for different SR that are predicted using a common control region are assumed to be correlated.

The experimental uncertainties related to the isolated electron, muon, and $\tauh$~vetoes, \bq~tagging, soft \bq~tagging, jet energy scale, \met resolution, reweighting for pileup, and top quark and \PW~boson tagging, are correlated across all SR and all backgrounds. The uncertainties in the lost lepton background estimate corresponding to differences in its \ttbar and $\PW$+jets fractions, or in the choice of PDF, $\alpha_S$, and $\mu_{\text{R}}/\mu_{\text{F}}$ are also correlated for all SR and the single-lepton CR. Uncertainties due to the lepton correction factors are treated as anticorrelated between the single-lepton CR and the SR, since an underestimate of the efficiency for selecting leptons in the CR corresponds to an overestimate in the efficiency for vetoing leptons in the SR. The uncertainties assigned to the \znunu~prediction are correlated separately to the uncertainties in $R_Z$ for all regions with the same $\nb$ (and same $\nsv$), and uncertainties originating from the discrepancies between the data-to-simulation ratios in $\zll+$jets and $\gamma+$jets events are correlated for all SR. The uncertainties in the QCD multijet background estimates corresponding to closure in the smearing method are correlated for all SR. For rare backgrounds, the uncertainties due to the variations of cross section, the PDF, $\alpha_S$, and $\mu_{\text{R}}/\mu_{\text{F}}$, are correlated for all SR.

For the simulated signal events, the differences between the fast simulation and the full \GEANTfour-based simulation are taken into account. Appropriate corrections and uncertainties for differences observed in lepton selection efficiencies, \bq~tagging performance, soft \bq~tagging performance, jet energy scale, and top quark and \PW~boson tagging efficiencies are applied to the predicted yields. The modeling of ISR plays an important role when it is relied upon to provide the $\stopq\stopqbar$ system with a significant momentum boost, such as in modeling low \dm~signals. To improve on the modeling of the multiplicity of additional jets from ISR, the signal events are reweighted based on the number of ISR jets ($N_\text{j}^{\text{ISR}}$) so as to make the jet multiplicity agree with data. The reweighting factors vary between 0.92 and 0.51 for $N_\text{j}^{\text{ISR}}$ between 1 and 6. We take one half of the deviation from unity as the systematic uncertainty on these reweighting factors. We also assess uncertainties in the predicted signal arising from changes in $\mu_{\text{R}}$ and $\mu_{\text{F}}$. An extra correction and uncertainty is assigned for possible differences in \met resolution between the fast and the full \GEANTfour-based simulations. This uncertainty is correlated among all SR under the assumption of a uniform prior.

\begin{figure}[!!ht]
\centering
\includegraphics[width=0.8\textwidth]{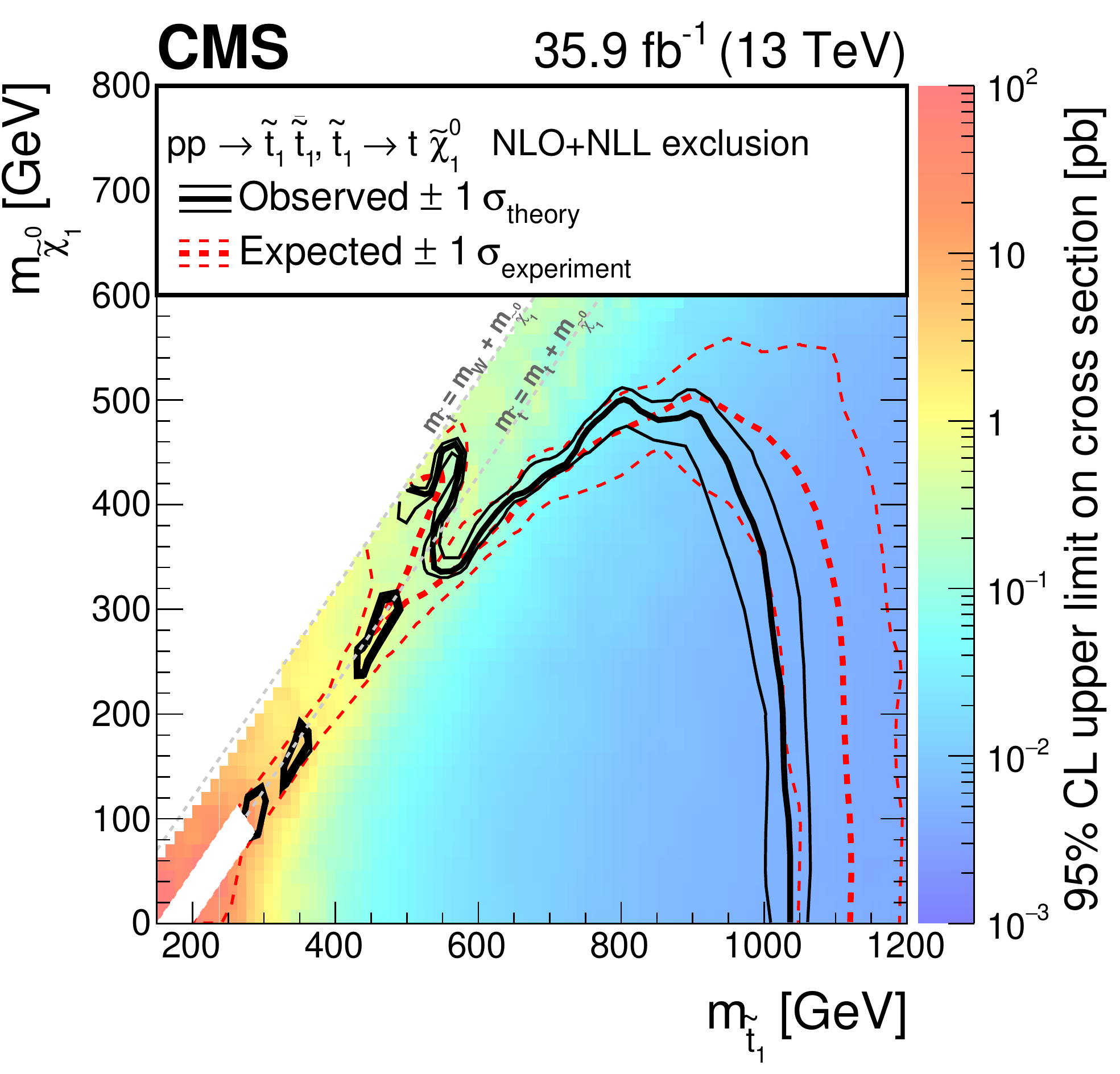}

\caption{\label{fig:T2ttlimits}Exclusion limits at 95\% CL for top squark pair production followed by the decay $\stopq \to \topq \lsp$ (T2tt), in the mass plane $m_{\lsp}$ versus $m_{\stopq}$. The areas to the left and below the solid black curves represent the observed exclusion and the $\pm$1 standard deviation contours for the NLO+NLL cross section calculations and their uncertainties~\cite{Borschensky:2014cia}. The dashed red curves represent the corresponding expectation at 95\% CL and $\pm$1 standard deviation contours for the associated experimental uncertainties. The ``islands" represent regions that are not excluded by this search. The diagonal area where \dm~is very close to the top quark mass, corresponding to a very light \lsp, is left blank because the final states, which are similar to the SM \ttbar~background, have acceptance that varies strongly with \lsp~mass, making it very difficult to model the signal acceptance in this region.}
\end{figure}

\begin{figure}[!!ht]
\centering
\includegraphics[width=0.8\textwidth]{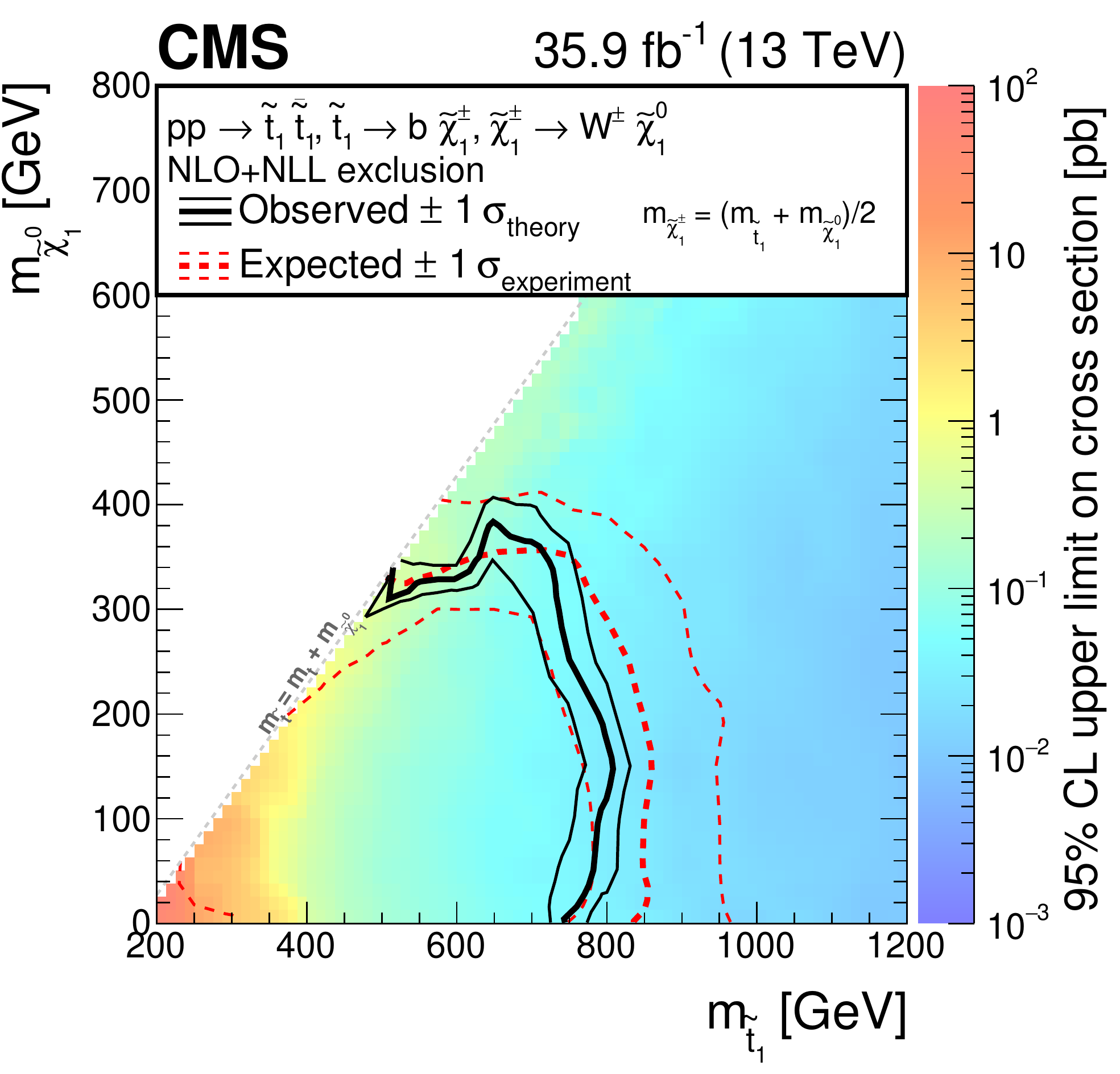}

\caption{\label{fig:T2bWlimits}Exclusion limits at 95\% CL for top squark pair production followed by the decay $\stopq \to \bq \chipmone \to \bq \PW^{\pm} \lsp$ (T2bW), in the mass plane $m_{\lsp}$ versus $m_{\stopq}$. The areas to the left and below the solid black curves represent the observed exclusion and the $\pm$1 standard deviation contours for the NLO+NLL cross section calculations and their uncertainties~\cite{Borschensky:2014cia}. The dashed red curves represent the corresponding expectation at 95\% CL and $\pm$1 standard deviation contours for the associated experimental uncertainties. In the lower left corner of the diagram, where \dm~is close to the top quark mass, the sensitivity of the search is significantly reduced due to the fact that the $\stopq$ decay products are soft and often escape detection. }
\end{figure}

\begin{figure}[!!ht]
\centering
\includegraphics[width=0.8\textwidth]{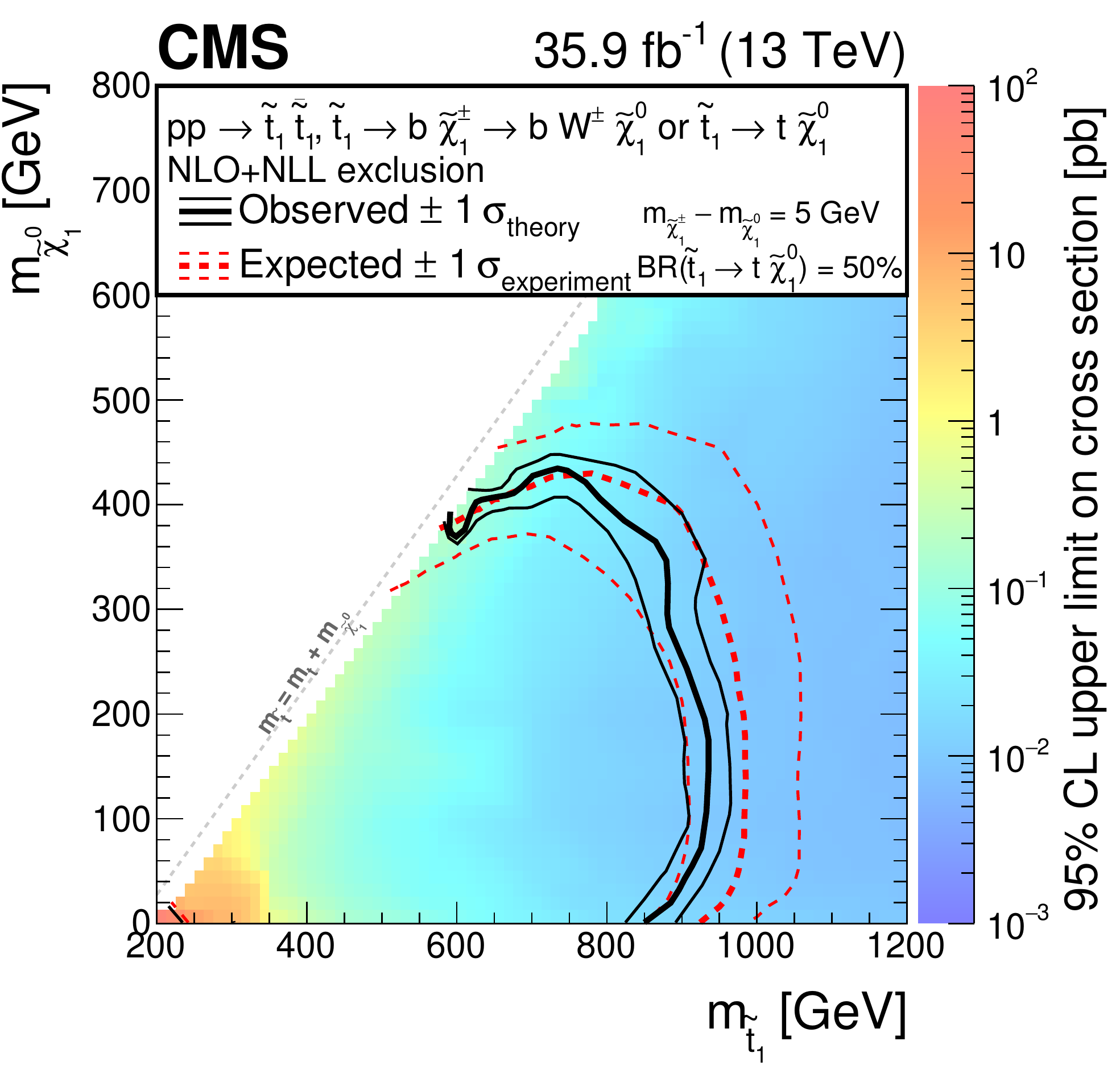}

\caption{\label{fig:T2tblimits}Exclusion limits at 95\% CL for top squark pair production followed by the mixed decay $\Pp\Pp\to\stopq\stopqbar\to \topqbar\lsp\bq\chgp$ (T2tb) decay scenario, in the mass plane $m_{\lsp}$ versus $m_{\stopq}$. In T2tb an assumption of a compressed mass spectrum in which the mass of \chipmone~is only 5\GeV greater than that of \lsp, is considered. The areas to the left and below the solid black curves represent the observed exclusion and the $\pm$1 standard deviation contours for the NLO+NLL cross section calculations and their uncertainties~\cite{Borschensky:2014cia}. The dashed red curves represent the corresponding expectation at 95\% CL and $\pm$1 standard deviation contours for the associated experimental uncertainties. In the lower left corner of the diagram, where \dm~is close to the top quark mass, the sensitivity of the search is significantly reduced due to the fact that the $\stopq$ decay products are soft and often escape detection.
}
\end{figure}

The results of the fit procedure are used to set exclusion limits in the models shown in Fig.~\ref{fig:diagram}. We set 95\% CL exclusion limits for the large \dm~signal models, namely the T2tt, T2bW, and T2tb models described in Section~\ref{sec:intro}, in the mass plane $m_{\lsp}$ versus $m_{\stopq}$, together with upper limits at 95\% CL on the signal cross section, under the hypotheses of the top squarks decaying as prescribed by these models. In the case of the T2tt model, Fig.~\ref{fig:T2ttlimits}, we can exclude $\stopq$ masses up to 1040\GeV and $\lsp$ masses up to 500\GeV. The blank region below the diagonal in the lower left corner of the figure corresponds to values of \dm~that are very close to the top quark mass when the \lsp is very light. This area is particularly challenging due to the similarity of the final states to SM \ttbar background, which results in a significant contamination from leptonic signal events in the control regions. Our analysis does not have adequate signal discrimination to provide exclusion limits in this region. Figure~\ref{fig:T2bWlimits} shows the exclusion limits obtained for the T2bW model. Under this decay hypothesis, for which the $\chipmone$ mass lies halfway between the $\stopq$ and $\lsp$ masses, we can exclude $\stopq$ masses up to 800\GeV and $\lsp$ masses up to 360\GeV. Figure~\ref{fig:T2tblimits} addresses the T2tb model, in which both of these decay modes are allowed with equal probability while also assuming a compressed mass spectrum in which the mass of the \chipmone~is only 5\GeV greater than that of the \lsp. We can exclude for this model $\stopq$ masses up to 940\GeV, and $\lsp$ masses up to 440\GeV.

\begin{figure}[!!ht]
\centering
\includegraphics[width=0.8\textwidth]{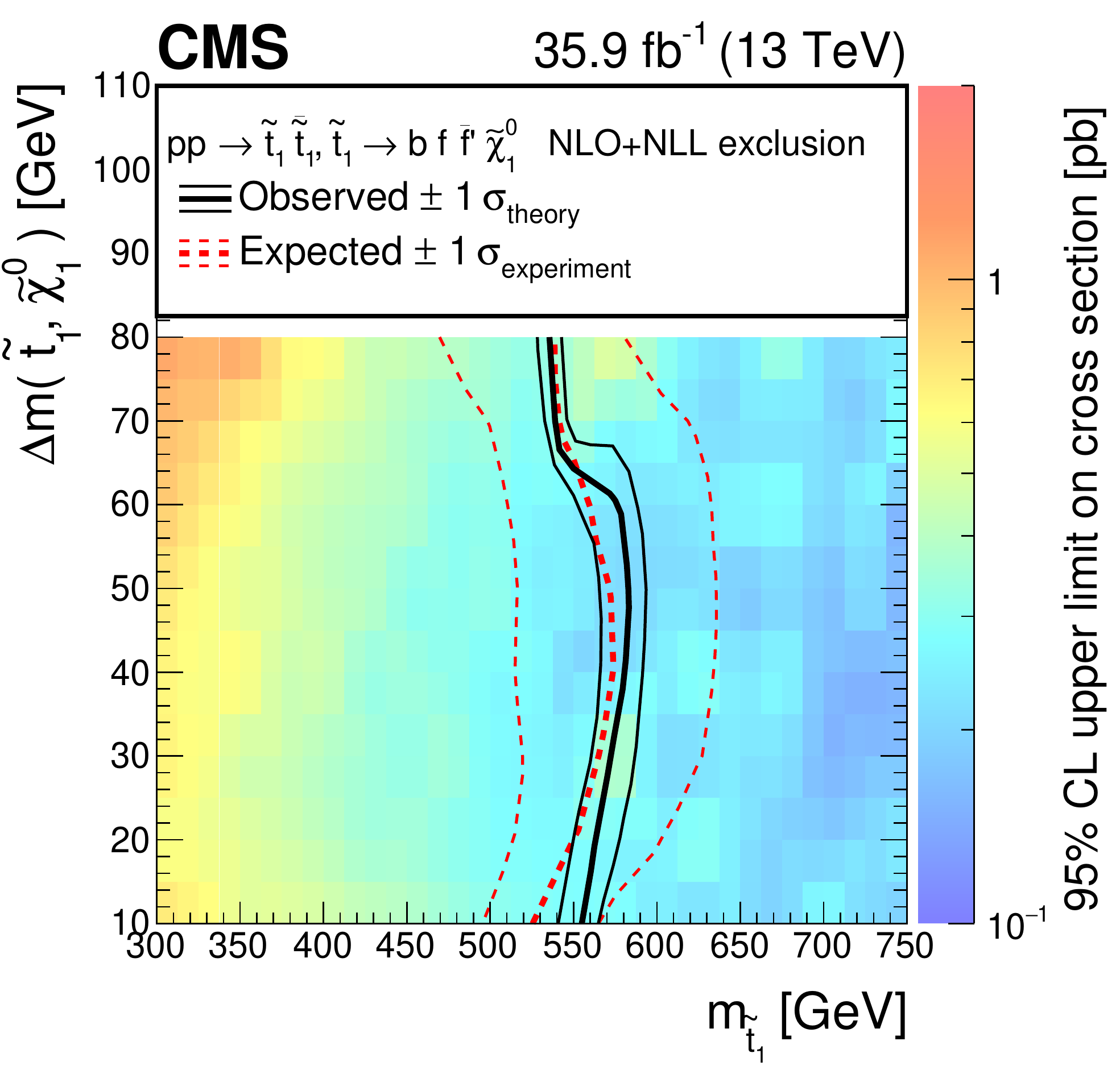}

\caption{\label{fig:T2ttClimits}Exclusion limits at 95\% CL for top squark pair production followed by the four-body decay $\stopq ~ \to~\bq\cmsSymbolFace{f} \bar{\cmsSymbolFace{f'}} \lsp$ (T2ttC) in the mass plane $\dm(\stopq,\lsp)$ versus $m_{\stopq}$. The areas to the left and below the solid black curves represent the observed exclusion and the $\pm$1 standard deviation contours for the NLO+NLL cross section calculations and their uncertainties~\cite{Borschensky:2014cia}. The dashed red curves represent the corresponding expectation at 95\% CL and $\pm$1 standard deviation contours for the associated experimental uncertainties. }
\end{figure}

\begin{figure}[!!ht]
\centering
\includegraphics[width=0.8\textwidth]{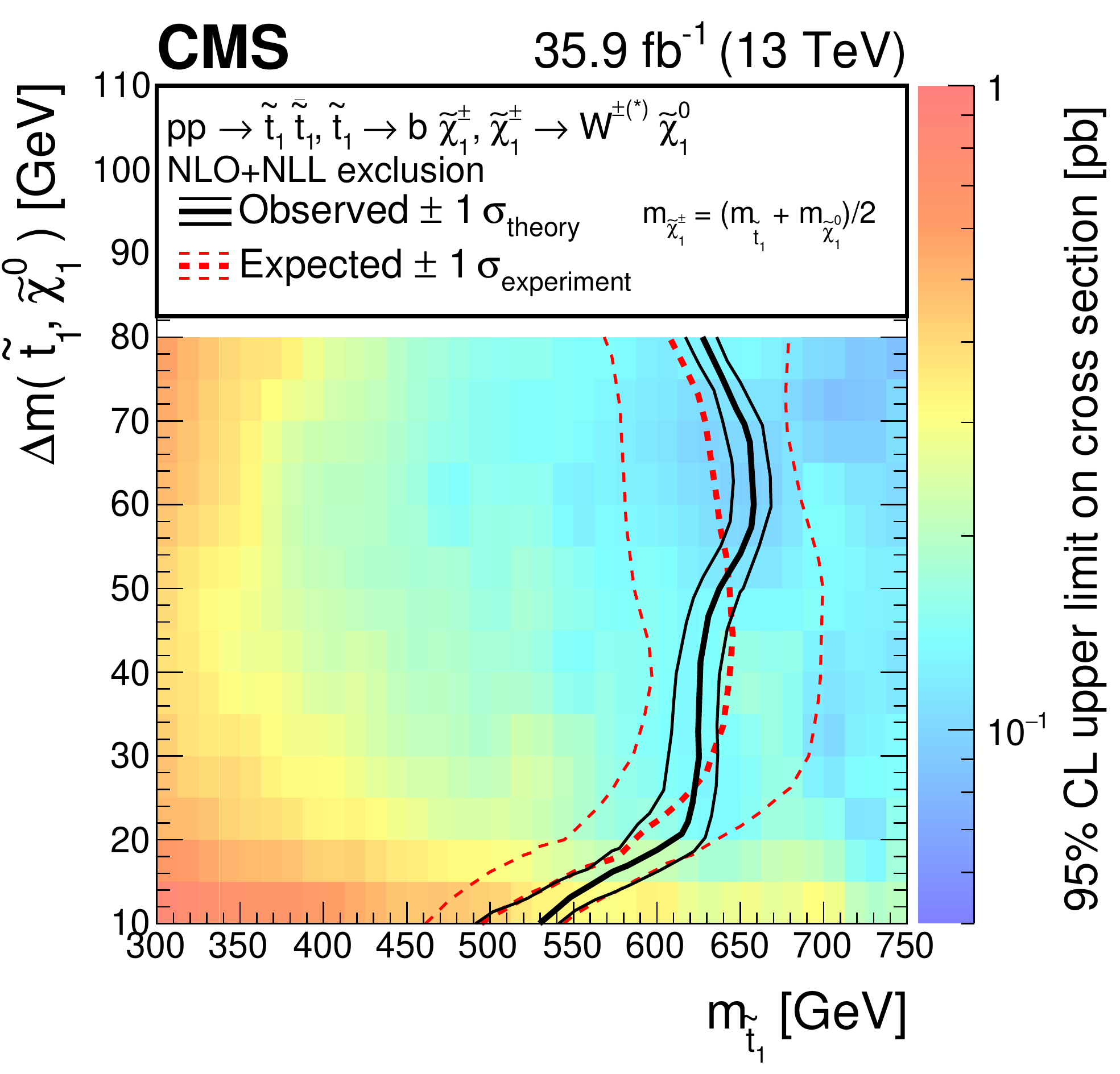}

\caption{\label{fig:T2bWClimits}Exclusion limits at 95\% CL for top squark pair production followed by the decay $\stopq \to \bq \chipmone \to \bq \cmsSymbolFace{f} \bar{\cmsSymbolFace{f'}} \lsp$ (T2bWC) in the mass plane $\dm(\stopq,\lsp)$ versus $m_{\stopq}$. The areas to the left and below the solid black curves represent the observed exclusion and the $\pm$1 standard deviation contours for the NLO+NLL cross section calculations and their uncertainties~\cite{Borschensky:2014cia}. The dashed red curves represent the corresponding expectation at 95\% CL and $\pm$1 standard deviation contours for the associated experimental uncertainties.
}
\end{figure}

\begin{figure}[!!ht]
\centering
\includegraphics[width=0.8\textwidth]{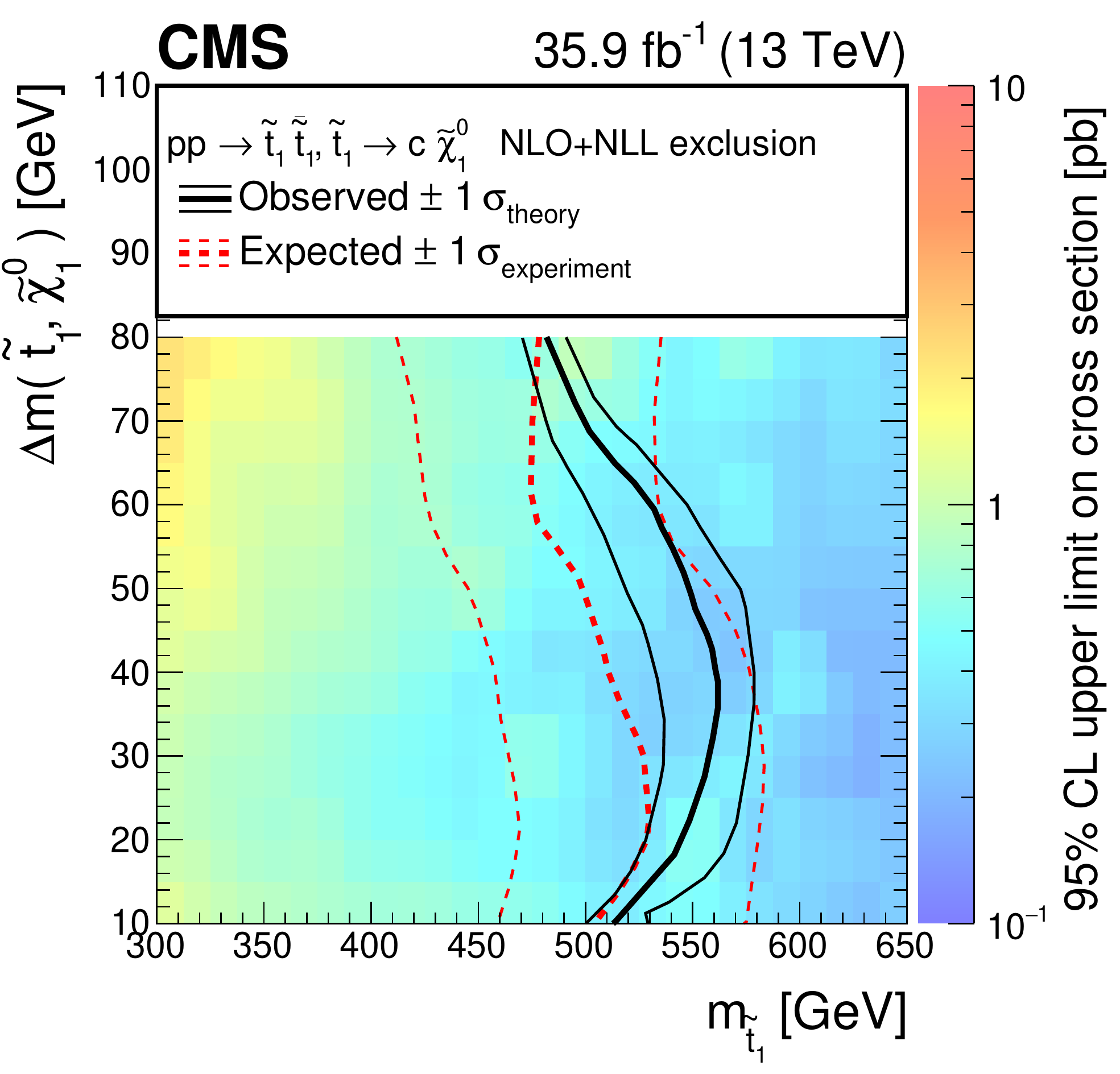}

\caption{\label{fig:T2cclimits}Exclusion limits at 95\% CL for top squark pair production followed by the decay $\stopq \to \cq\lsp$ (T2cc) in the mass plane $\dm(\stopq,\lsp)$ versus $m_{\stopq}$. The areas to the left and below the solid black curves represent the observed exclusion and the $\pm$1 standard deviation contours for the NLO+NLL cross section calculations and their uncertainties~\cite{Borschensky:2014cia}. The dashed red curves represent the corresponding expectation at 95\% CL and $\pm$1 standard deviation contours for the associated experimental uncertainties.
}
\end{figure}

We also set exclusion limits in the mass plane \dm~versus $m_{\stopq}$ for small \dm~signal models in which the mass difference between the $\stopq$ and $\lsp$ is smaller than the \PW~boson mass, namely the T2ttC, T2bWC, and T2cc models  described in Section~\ref{sec:intro}. Note that for mass configurations with $\dm < 30\GeV$,  the $\stopq$~lifetime becomes significant \cite{Grober:2014aha}. This is not taken into account in the simulation that we used for the T2ttC model. We therefore also consider the T2bWC model, which ensures reasonable lifetimes for the $\stopq$ in this region where the~\chipmone decays to an off-shell \PW~boson and an \lsp. Figures~\ref{fig:T2ttClimits} and \ref{fig:T2bWClimits} show the exclusion limits obtained for exclusive T2ttC and exclusive T2bWC decays, respectively. We can exclude $\stopq$ masses up to 580 (660)\GeV, and $\lsp$ masses up to 540 (610)\GeV for the T2ttC (T2bWC) model. Finally, Fig.~\ref{fig:T2cclimits}~shows the exclusion limits obtained for the T2cc model in which we exclude $\stopq$ and $\lsp$ masses up to 560 and 520\GeV, respectively.

\section{Summary}
A search is presented for direct top squark pair production in the all-jets final states based upon data collected with the CMS detector in \Pp\Pp~collisions at a center-of-mass energy of 13\TeV, corresponding to an integrated luminosity of 35.9\fbinv. The search is optimized for discovery through a variety of signatures. No significant excess of events is observed beyond the expected contribution from SM processes, and exclusion limits are set at a 95\% confidence level in the context of simplified models \cite{Alwall:2008ag,Alwall:2008va,Alves:2011wf} of direct top squark pair production.

In the parameter space of large mass differences between the $\stopq$ and $\lsp$ that permit the $\stopq$ to decay to an on-shell top quark and a neutralino, top squark masses up to 1040\GeV and $\lsp$ masses up to 500\GeV are excluded. Alternatively, when the top squark decays to a bottom quark and a $\chipmone$, $\stopq$ masses up to 800\GeV and $\lsp$ masses up to 360\GeV are excluded. Finally, for possibilities in which the branching fractions for these two top squark decay modes equal 50\%, including the assumption of a compressed mass spectrum with the mass of the \chipmone~ only 5\GeV greater than that of \lsp, top squark masses up to 940\GeV and $\lsp$ masses up to 440\GeV are excluded.

In the regions of parameter space where the mass difference between the $\stopq$ and $\lsp$ is smaller than the mass of the \PW~boson, we consider four-body decays of top squarks in which top squark masses up to 580\GeV are excluded for a neutralino mass of 540\GeV. An additional decay that is relevant in this parameter space is one in which the top squark decays to a bottom quark and a $\chipmone$, that then decays to a virtual \PW~boson and a \lsp. Here, top squark masses up to 660\GeV are excluded for a neutralino mass of 610\GeV. Finally, we consider decays through a flavor changing neutral current process where the $\stopq$ decays to a \cq~quark and a \lsp. In this case, $\stopq$~and \lsp~masses up to 560\GeV~and up to 520\GeV, respectively, are excluded.

In summary, we present a search that takes advantage of a large new set of data collected by the CMS experiment in 2016, as well as a variety of new methods that yield exclusion limits for a wide array of top squark decay modes in planes of $m_{\lsp}$ versus $m_{\stopq}$ and  $m_{\stopq} - m_{\lsp}$ versus $m_{\stopq}$ that extend significantly beyond those obtained in previous searches.

\begin{acknowledgments}
We congratulate our colleagues in the CERN accelerator departments for the excellent performance of the LHC and thank the technical and administrative staffs at CERN and at other CMS institutes for their contributions to the success of the CMS effort. In addition, we gratefully acknowledge the computing centers and personnel of the Worldwide LHC Computing Grid for delivering so effectively the computing infrastructure essential to our analyses. Finally, we acknowledge the enduring support for the construction and operation of the LHC and the CMS detector provided by the following funding agencies: BMWFW and FWF (Austria); FNRS and FWO (Belgium); CNPq, CAPES, FAPERJ, and FAPESP (Brazil); MES (Bulgaria); CERN; CAS, MoST, and NSFC (China); COLCIENCIAS (Colombia); MSES and CSF (Croatia); RPF (Cyprus); SENESCYT (Ecuador); MoER, ERC IUT, and ERDF (Estonia); Academy of Finland, MEC, and HIP (Finland); CEA and CNRS/IN2P3 (France); BMBF, DFG, and HGF (Germany); GSRT (Greece); OTKA and NIH (Hungary); DAE and DST (India); IPM (Iran); SFI (Ireland); INFN (Italy); MSIP and NRF (Republic of Korea); LAS (Lithuania); MOE and UM (Malaysia); BUAP, CINVESTAV, CONACYT, LNS, SEP, and UASLP-FAI (Mexico); MBIE (New Zealand); PAEC (Pakistan); MSHE and NSC (Poland); FCT (Portugal); JINR (Dubna); MON, RosAtom, RAS, RFBR and RAEP (Russia); MESTD (Serbia); SEIDI, CPAN, PCTI and FEDER (Spain); Swiss Funding Agencies (Switzerland); MST (Taipei); ThEPCenter, IPST, STAR, and NSTDA (Thailand); TUBITAK and TAEK (Turkey); NASU and SFFR (Ukraine); STFC (United Kingdom); DOE and NSF (USA).

\hyphenation{Rachada-pisek} Individuals have received support from the Marie-Curie program and the European Research Council and Horizon 2020 Grant, contract No. 675440 (European Union); the Leventis Foundation; the A. P. Sloan Foundation; the Alexander von Humboldt Foundation; the Belgian Federal Science Policy Office; the Fonds pour la Formation \`a la Recherche dans l'Industrie et dans l'Agriculture (FRIA-Belgium); the Agentschap voor Innovatie door Wetenschap en Technologie (IWT-Belgium); the Ministry of Education, Youth and Sports (MEYS) of the Czech Republic; the Council of Science and Industrial Research, India; the HOMING PLUS program of the Foundation for Polish Science, cofinanced from European Union, Regional Development Fund, the Mobility Plus program of the Ministry of Science and Higher Education, the National Science Center (Poland), contracts Harmonia 2014/14/M/ST2/00428, Opus 2014/13/B/ST2/02543, 2014/15/B/ST2/03998, and 2015/19/B/ST2/02861, Sonata-bis 2012/07/E/ST2/01406; the National Priorities Research Program by Qatar National Research Fund; the Programa Clar\'in-COFUND del Principado de Asturias; the Thalis and Aristeia programs cofinanced by EU-ESF and the Greek NSRF; the Rachadapisek Sompot Fund for Postdoctoral Fellowship, Chulalongkorn University and the Chulalongkorn Academic into Its 2nd Century Project Advancement Project (Thailand); and the Welch Foundation, contract C-1845. \end{acknowledgments}
\clearpage
\bibliography{auto_generated}

\cleardoublepage \appendix\section{The CMS Collaboration \label{app:collab}}\begin{sloppypar}\hyphenpenalty=5000\widowpenalty=500\clubpenalty=5000\input{SUS-16-049-authorlist.tex}\end{sloppypar}
\end{document}

%% file: SUS-16-049-authorlist.tex
\textbf{Yerevan Physics Institute,  Yerevan,  Armenia}\\*[0pt]
A.M.~Sirunyan, A.~Tumasyan
\vskip\cmsinstskip
\textbf{Institut f\"{u}r Hochenergiephysik,  Wien,  Austria}\\*[0pt]
W.~Adam, F.~Ambrogi, E.~Asilar, T.~Bergauer, J.~Brandstetter, E.~Brondolin, M.~Dragicevic, J.~Er\"{o}, M.~Flechl, M.~Friedl, R.~Fr\"{u}hwirth\cmsAuthorMark{1}, V.M.~Ghete, J.~Grossmann, J.~Hrubec, M.~Jeitler\cmsAuthorMark{1}, A.~K\"{o}nig, N.~Krammer, I.~Kr\"{a}tschmer, D.~Liko, T.~Madlener, I.~Mikulec, E.~Pree, D.~Rabady, N.~Rad, H.~Rohringer, J.~Schieck\cmsAuthorMark{1}, R.~Sch\"{o}fbeck, M.~Spanring, D.~Spitzbart, J.~Strauss, W.~Waltenberger, J.~Wittmann, C.-E.~Wulz\cmsAuthorMark{1}, M.~Zarucki
% ...
\vskip\cmsinstskip
% ...
1:~~Also at Vienna University of Technology, Vienna, Austria\\
% ...
